\documentclass[
    aps,
    prd,
    notitlepage,
    twocolumn,s
    tightenlines,
    nofootinbib,
    superscriptaddress]{revtex4-2}
\pdfoutput=-1
\usepackage[linktocpage,breaklinks]{hyperref}
\usepackage[usenames,dvipsnames]{xcolor}
\usepackage{amsmath}
\usepackage{amssymb}
\usepackage{mathtools}
\usepackage{amsfonts}
\usepackage{txfonts}
\usepackage{bm}
\usepackage{stmaryrd}
\usepackage{tensor}
\usepackage{mathrsfs}
\usepackage[utf8]{inputenc}
\usepackage{url}
\usepackage{graphicx}
\usepackage{dcolumn}
\usepackage{epsfig}
\usepackage{epstopdf}
\usepackage[normalem]{ulem}
\usepackage{natbib}
\usepackage{hyperref}
\allowdisplaybreaks
\usepackage{stackengine}
\usepackage{cleveref}
\hypersetup{colorlinks=true,
            citecolor=Black,
            linkcolor=Black,
            urlcolor=Black}

\definecolor{darkred}{rgb}{0.8, 0.0, 0.0}

\begin{document}
\title{Petrov Type, Principal Null Directions, and Killing Tensors of\\Slowly-Rotating Black Holes in Quadratic Gravity}
\author{Caroline B. Owen 
}
\email{cbo4@illinois.edu}
\affiliation{%
Illinois Center for Advanced Studies of the Universe, Department of Physics,
University of Illinois at Urbana-Champaign, Urbana, IL 61801, USA}%

\author{Nicol\'as Yunes}
\email{nyunes@illinois.edu}
\affiliation{%
Illinois Center for Advanced Studies of the Universe, Department of Physics,
University of Illinois at Urbana-Champaign, Urbana, IL 61801, USA}%

\author{Helvi Witek}
\email{hwitek@illinois.edu}
\affiliation{%
Illinois Center for Advanced Studies of the Universe, Department of Physics,
University of Illinois at Urbana-Champaign, Urbana, IL 61801, USA}%

\date{July 23, 2021}

\begin{abstract}
The ability to test general relativity in extreme gravity regimes using gravitational wave observations from current ground-based or future space-based detectors motivates the mathematical study of the symmetries of black holes in modified theories of gravity. In this paper we focus on spinning black hole solutions in two quadratic gravity theories: dynamical Chern-Simons and scalar Gauss-Bonnet gravity. We compute the 
principal null directions,
Weyl scalars, and complex null tetrad in the small-coupling, slow rotation approximation for both theories, confirming that both spacetimes are Petrov type I. Additionally, we solve the Killing equation through rank 6 in dynamical Chern-Simons gravity and rank 2 in scalar Gauss-Bonnet gravity, showing that there is no nontrivial Killing tensor through those ranks for each theory. We therefore conjecture that the still-unknown, exact, quadratic-gravity, black-hole solutions do not possess a fourth constant of motion.   
\end{abstract}
\maketitle
\section{Introduction}
Gravitational wave physics offers a novel way of studying cataclysmic astronomical events light years away, potentially elucidating the birth of our universe, or testing and refining our theory of gravity itself. It is this last option that interests us here. General relativity (GR) is our current best theory of gravity because it has passed a multitude of tests in weak gravity regimes, such as the solar system \cite{Will:2014kxa} and with binary pulsars \cite{Stairs:2003eg}. However, there are many anomalies, such as observations of galaxy rotation curves or the late-time acceleration of the universe, that GR cannot explain without the inclusion of certain ``dark'' components. Considering GR's successful track record in weak gravitational systems, it stands to reason that, if new physics is to be discovered to address these or other anomalies, it may be found through observations of extreme gravity systems, where the curvature is both large and dynamical~\cite{Yunes:2013dva,Berti:2015itd,Yunes:2016jcc}. 

A quintessential observation of this type is the gravitational waves emitted in compact binary coalescence~\cite{Yunes:2013dva}. Such observations require a waveform model for the gravitational waves emitted in the inspiral, merger, and ringdown.
The models rely on an understanding of the binary dynamics, which in turn, first necessitates an understanding of isolated black hole (BH) solutions. In GR, this understanding was developed during the 1960s golden age, when exact solutions representing spinning and charged black holes were discovered \cite{Kerr_1963,Newman:1965my}, their symmetry properties analyzed~\cite{Hawking:1971tu,Hawking:1971vc}, no hair theorems proven~\cite{Israel:1967wq,Israel:1967za,Carter:1971zc}, and stability properties understood through the evolution of their perturbations \cite{Press:1973zz}. Such an analysis is lacking in most, if not all, BH solutions found in modified theories of gravity. This gap is in part due to a lack of closed-form exact solutions for most modified theories, with only approximate solutions known when BHs spin slowly. Therefore, tests of GR that involve such modified gravity BHs may soon necessitate a more thorough understanding of their mathematical structure. 

BHs in modified gravity may be drastically different from those in GR, even while recovering GR predictions when expanded in the far field and to leading order. One area where this is very clear is in their Petrov classification \cite{Petrov:2000bs}, i.e.~the BHs of GR are Petrov type D \cite{Chandrasekhar_1992}, while those of modified gravity need not be \cite{Araneda_2015}. A classification of type D comes with many benefits that simplify calculations and make difficult problems more tractable, such as when studying extreme mass-ratio inspirals~\cite{AmaroSeoane:2010zy}. One such benefit is the ability to choose a tetrad frame in the Newman-Penrose formalism in which all but one of the Weyl scalars vanish and two vectors of the tetrad are aligned with the principal null directions (PNDs) of the spacetime. The ability to choose such a frame was an essential assumption in the derivation of Teukolsky's equations, which describe the evolution of BH perturbations and characterize their stability \cite{Teukolsky:1973ha}. 

Another remarkable property of Petrov type D vacuum solutions is the guaranteed existence of a rank-2 Killing tensor. Killing tensors relate to symmetries of the spacetime and, when contracted completely with copies of a geodesic four-velocity, generate scalar quantities that are conserved along said geodesic. For the Kerr metric of GR, this rank-2 Killing tensor generates the Carter constant, which, along with energy, a component of angular momentum, and rest mass, brings the number of conserved quantities to 4 for a test particle moving along a geodesic \cite{Carter:1968rr,Walker:1970un}. When the number of conserved quantities equals the number of degrees of freedom in a system, the equations of motion can be cast in quadrature form, and when this is not possible, geodesic motion is said to be Liouville chaotic \cite{Schuster_2006}.

Modified gravity BHs, on the other hand, are not guaranteed to be of Petrov type D, and as a result, all of the nice results described above do not extend naturally beyond GR. We are therefore motivated to study BHs with a generic mathematical framework that does not rely on a type D classification. In this paper, we take steps toward that goal. We focus on the spinning BH solutions of two modified theories: dynamical Chern-Simons gravity \cite{Alexander:2009tp} and scalar Gauss-Bonnet gravity \cite{Kanti:1995vq}, both of which are classified as quadratic gravity theories~\cite{Yunes_2011}. We present the PNDs, complex null tetrad, and Weyl scalars in the small-coupling, slow-rotating limit for the BH solutions in both theories. In each case, we find that the leading order corrections to the PNDs cause each of the GR PNDs to split in two, confirming that both solutions are indeed not Petrov type D but instead Petrov type I. We begin by constructing an orthonormal tetrad from each metric and use it to compute a complex null tetrad, following the conventions of the Newman-Penrose formalism. We use this tetrad to determine the Petrov type and to construct the PNDs. We then transform the tetrad and Weyl scalars into a frame where one vector is a PND, the Weyl scalars $\Psi_0$ and $\Psi_4$ vanish, and the GR parts match what is standard in the literature. 

We also explore the existence of a fourth constant of motion in each theory by solving the Killing equation perturbatively in spin and coupling. Previous work has shown that while the spinning BH of dynamical Chern-Simons (dCS) gravity does possess a rank-2 Killing tensor that leads to a Carter-like constant at linear order in spin~\cite{Sopuerta_2011}, there is no extension of this Killing tensor at quadratic order in spin~\cite{Yagi_2012}. Here, we expand those results significantly to show that dCS BHs do not possess a nontrivial Killing tensor through rank 6. We therefore conjecture that the yet unknown exact BH solution of dCS gravity does not possess a fourth constant of motion. We also compute a rank-2 Killing tensor for scalar Gauss-Bonnet (sGB) gravity through linear order in spin and show, just as in dCS gravity, that this Killing tensor cannot be extended to quadratic order in spin.

The remainder of this paper presents the details of the results summarized above. Section \ref{SEC:Black Holes in Quadratic Gravity} gives an overview of quadratic gravity as well as details about dCS and sGB gravity and their spinning BH solutions. Section \ref{sec:Petrov Type, Principal Null Directions, and Newman-Penrose Formalism} presents a method to determine the Petrov type and compute the PNDs, complex null tetrad, and Weyl scalars in a broad class of stationary, axially symmetric black holes, and applies the method to the spinning BHs of dCS and sGB gravity. Section \ref{SEC:Conserved Quantities}  investigates the existence of a fourth constant of motion in both theories. Finally, Sec.~\ref{SEC:Conclusion} summarizes and details future avenues of research that could build on our results. In the following, we use geometric units $G=1=c$ and the $(-,+,+,+)$ metric signature. Complex conjugation is indicated with an overbar; i.e. the complex conjugate of $A$ is $\bar{A}$. Symmetrization over indices is denoted with parentheses, such that $A_{(\alpha\beta)} = \frac{1}{2}\left(A_{\alpha\beta} + A_{\beta\alpha}\right)$.

\section{Black Holes in Quadratic Gravity}\label{SEC:Black Holes in Quadratic Gravity}
Quadratic gravity is a class of modified theories of gravity that introduce a scalar field $\vartheta$ coupled to the gravitational field through quadratic curvature invariants. Such theories have been studied intensely in recent years (see e.g.~\cite{Yunes_2011} for a recent review). We here present the basics again to establish notation, following mostly the presentation in \cite{Yagi:2015oca}. In these theories, the action takes the form
\begin{align}
    S = S_\mathrm{EH} + S_\mathrm{mat} + S_\vartheta + S_q,
\end{align}
which contains the 
Einstein-Hilbert term of GR,  
\begin{align}
    S_\mathrm{EH} =&~ \kappa\int_\mathcal{V} d^4x\sqrt{-g} R, 
\end{align}
where $\kappa = (16 \pi G)^{-1}$ and $g$ is the determinant of the metric $g_{\mu\nu}$, 
a matter term $S_\mathrm{mat}$ that depends only on the metric and matter fields, but not the scalar field, and the canonical scalar field action 
\begin{align}
 S_\vartheta =&~ - \frac{\beta}{2}\int_\mathcal{V} d^4x\sqrt{-g}
 \left[\nabla_\mu\vartheta\nabla^\mu\vartheta + 2V(\vartheta)\right],
\end{align}
where  both the scalar field $\vartheta$ and the parameter $\beta$ are taken to be dimensionless.
In the following we set $V(\vartheta)=0$ because we are interested in massless fields. 
 The quadratic term $S_q$ prescribes the coupling of $\vartheta$ to the independent quadratic curvature invariants $R^2$, $R_{\alpha\beta}R^{\alpha\beta}$, the Kretschmann scalar $R_{\alpha\beta\gamma\delta}R^{\alpha\beta\gamma\delta}$, and the Pontryagin density $R_{\alpha\beta\gamma\delta}^{}\,^{*}R^{\alpha\beta\gamma\delta}$. These invariants are constructed from the Ricci scalar  $R = g^{\alpha\beta}g^{\gamma\delta}R_{\gamma\alpha\delta\beta}$, the Ricci tensor $R_{\alpha\beta}=g^{\gamma\delta}R_{\gamma\alpha\delta\beta}$, the Riemann tensor,  $R_{\gamma\alpha\delta\beta}$ and its dual $^{*}R_{\alpha\beta\delta\gamma} = \frac{1}{2}\epsilon_{\alpha\beta}^{}{}^{\mu\nu}R_{\mu\nu\gamma\delta}$.

Quadratic gravity theories find motivation in several places. One way such theories can be motivated is by thinking of GR as an effective field theory valid for small curvatures \cite{Yunes_2011}. In this context, the Einstein-Hilbert action can be seen as the first term in an expansion in powers of the Riemann curvature tensor. The quadratic action $S_q$ would then constitute the next term in the expansion, necessary for more accurately describing phenomena in extreme gravity systems. Quadratic gravity also finds motivation in theories of quantum gravity that introduce higher curvature corrections and scalar fields to GR \cite{Alexander_2006,Taveras_2008,Weinberg_2008,Boulware_1985,
Kanti:1995vq,Alexander:2009tp}.

Black holes in quadratic gravity typically carry scalar or axion ``hair''~\cite{Campbell:1990ai,Yunes_2009,Kanti:1995vq,Sotiriou_2014} that yields scalar dipole radiation when the BH is placed in a compact binary~\cite{Yagi:2015oca}. This phenomenon is interesting because such radiation would accelerate the rate at which compact binaries inspiral~\cite{Yagi:2011xp,Okounkova:2017yby,Witek:2018dmd,East:2020hgw,Okounkova:2020rqw,
Silva:2020omi}.  The acceleration would, in turn, imprint in the gravitational wave phase~\cite{Shiralilou:2020gah} to which  interferometers such as LIGO~\cite{TheLIGOScientific:2014jea}, Virgo~\cite{TheVirgo:2014hva} and KAGRA~\cite{Akutsu:2020his} are most sensitive.
The presence of such an enhanced rate of inspiral would be smoking-gun evidence for a deviation of GR, while the absence could help stringently constrain quadratic gravity theories~\cite{Perkins:2020tra,Nair_2019}. These tests, however, may require more accurate waveform models, especially when considering extreme mass-ratio inspirals, and this, in turn, necessitates a deep understanding of the mathematical structure of quadratic gravity BHs.   

In this paper we consider two quadratic gravity theories: dynamical Chern-Simons gravity and scalar Gauss-Bonnet gravity. We go into further detail on both theories in the following subsections.
\subsection{Dynamical Chern-Simons Gravity}
Dynamical Chern-Simons gravity modifies GR through \cite{Alexander:2009tp}
\begin{align}
 S_{q,\mathrm{CS}} =&~ \frac{\alpha_\mathrm{CS}}{4}\int_\mathcal{V} d^4x\sqrt{-g}\vartheta R_{\mu\nu\rho\sigma}^{}{}^{*}R^{\mu\nu\rho\sigma},
\end{align}
which couples a pseudoscalar $\vartheta$ field to the parity odd Pontryagin density. This correction to GR serves as a means to parametrize gravitational parity violation and finds motivation in string theory \cite{Alexander_2006}, loop quantum gravity \cite{Taveras_2008}, and inflation \cite{Weinberg_2008}. The coupling parameter $\alpha_\mathrm{CS}$ has dimensions of length squared and has been constrained to $\alpha_\mathrm{CS}^{1/2} \le 8.5 \mathrm{km}$ with $90\%$ confidence
using mass and equatorial plane measurements of an isolated neutron star~\cite{Silva_2020}.

The Pontryagin density vanishes for spherically symmetric spacetimes, so the Schwarzschild metric of GR is also a solution in dCS gravity. The Kerr metric of GR, on the other hand, is not a solution because the Pontryagin density sources a nontrivial scalar field yielding ``hairy'' black hole solutions. Currently, there is not an exact \textit{closed-form} solution describing spinning BHs in dCS gravity, although numerical solutions~\cite{Sullivan:2020zpf} and  small-coupling approximate solutions in both the slow-rotation \cite{Yunes_2009,Pani_2011,Yagi_2012,Maselli_2017}
and extremal \cite{McNees_2016} regimes do exist.  

In the small coupling approximation, deformations from the Kerr metric in Boyer-Lindquist coordinates $(t,r,\theta,\phi)$
\begin{align}
\label{EQ:Kerrmetric}
 ds^2_\mathrm{GR} = &~- \left(1 - \frac{2 M  r}{\rho^2}\right)dt^2 - \frac{4 M a r \sin^2\theta}{\rho^2}dtd\phi\nonumber\\
&~+\frac{\Sigma}{\rho^2}\sin^2\theta d\phi^2 + \frac{\rho^2}{\Delta}dr^2 + \rho^2d\theta^2,   
\end{align}
with metric functions
\begin{align}
\rho^2 = &~ r^2 + a^2\cos\theta^2,\quad
\Delta =  ~ r^2 - 2 M r + a^2,\nonumber\\
\Sigma = &~ (r^2+a^2)^2 - a^2\Delta \sin^2\theta,
\label{EQ:Kerrmetricfunctions}
\end{align}
are proportional to the dimensionless coupling parameter
\begin{align}
   \zeta \equiv \frac{\alpha_{CS}^2}{\kappa \beta M^4},
\end{align}
which is taken to be much less than unity $\zeta \ll 1$. In this paper, we focus on slowly rotating and small-coupling approximate solutions,
in which both the GR part of the metric and the dCS correction are additionally expanded in the dimensionless spin parameter $\chi = a/M \ll 1. $ Since the Schwarzschild metric is a solution in dCS gravity, there is no $\mathcal{O}(\chi^0\zeta)$ term in the metric and the leading-order-in-spin dCS correction is of $\mathcal{O}(\chi\zeta)$. The metric in the small-coupling, slow-rotation regime, known to $\mathcal{O}(\chi^5\zeta)$, is included in Appendix \ref{SEC:dCSMetric} for completeness \cite{Yunes_2009,Yagi_2012,Maselli_2017}.
\subsection{Scalar Gauss-Bonnet Gravity}
Scalar Gauss-Bonnet gravity is derived from the quadratic action~\cite{Kanti:1995vq}
\begin{align}
S_{q,\mathrm{GB}} = \alpha_\mathrm{GB}\int_\mathcal{V} d^4x\sqrt{-g}f(\vartheta) \mathcal{G},
\end{align}
where 
\begin{align}
\mathcal{G} = R^2 - 4R_{\alpha\beta}R^{\alpha\beta} + R_{\alpha\beta\gamma\delta}R^{\alpha\beta\gamma\delta}
\end{align}
is the parity even Gauss-Bonnet 
invariant. We consider coupling functions $f(\vartheta)$ that admit a Taylor expansion $f(\vartheta) = f(0) + f'(0)\vartheta +\mathcal{O}(\vartheta^2)$ about small $\vartheta$
with $f'(0)\neq0$.
Because $\mathcal{G}$ is a topological invariant, the first term in the expansion leads to a theory identical to GR and can be disregarded. We therefore focus on linear coupling functions $f(\vartheta) = f'(0)\vartheta$ and absorb the coefficient $f'(0)$ into the coupling parameter so $\alpha_{GB} f(\vartheta) \rightarrow \alpha_{GB} \vartheta$. Scalar Gauss-Bonnet gravity finds motivation in low-energy expansions of string theory \cite{Boulware_1985,Kanti:1995vq} and has been studied extensively \cite{Yunes_2011,Sotiriou_2014,Ripley:2019aqj,Benkel:2016rlz}.
The sGB coupling parameter $\alpha_{GB}$ has dimensions of length squared and has been constrained to $\alpha_{GB}^{1/2} \le 5.6\mathrm{km}$ with 90\% confidence using observations of coalescing binary black holes by the LIGO/Virgo Collaboration \cite{Nair_2019}.

As with dCS gravity, we work in the small-coupling, slow rotation approximation of sGB gravity \cite{Yunes_2011,Pani_2011,Ayzenberg_2014,Maselli:2015tta}. In the small-coupling regime, deformations from GR are proportional to the dimensionless coupling parameter
\begin{align}
   \zeta \equiv \frac{\alpha_{GB}^2}{\kappa \beta M^4}.
\end{align}
The dimensionless coupling parameter of sGB gravity is not the same as the dimensionless coupling parameter of dCS gravity, but to keep notation simple, we have called them both $\zeta$. As we never consider both theories simultaneously, it should always be clear to which we are referring.

Unlike the Pontryagin density, the Gauss-Bonnet 
invariant does not vanish for spherically symmetric spacetimes and, so long as $f'(\vartheta)\ne0$, the Schwarzschild metric is not a solution of sGB gravity.
Therefore, in contrast to dCS gravity, there is an $\mathcal{O}(\chi^0\zeta)$ term in the metric of sGB gravity. The small-coupling, slow-rotation sGB metric is known to $\mathcal{O}(\chi^5\zeta^7)$, but to parallel our calculation in dCS gravity we will be working with the sGB BH metric to $\mathcal{O}(\chi^5\zeta)$. The metric is included in Appendix \ref{SEC:dCSMetric} for completeness \cite{Yunes_2011,Pani_2011,Ayzenberg_2014,Maselli:2015tta}.
\section{\label{sec:Petrov Type, Principal Null Directions, and Newman-Penrose Formalism} Petrov Type, Principal Null Directions, and Newman-Penrose Formalism}

In this section, we present a method for constructing the complex null tetrad of the Newman-Penrose formalism in a particular frame for a 
broad class of stationary, axially symmetric spacetimes. We then demonstrate how to use the tetrad to determine the Petrov type of such a spacetime and construct its principal null directions. Additionally, we outline how to rotate the complex null tetrad and Weyl scalars into the conventional choice of frame for BHs in GR. The relevant quantities are presented in the small coupling limit for the slowly rotating BHs of dCS and sGB gravity. 

\subsection{\label{SEC:Basics in GR}Basics in GR}
The BH solutions we are interested in are stationary, axially symmetric, asymptotically flat vacuum solutions. A broad class of spacetimes with these properties can be described with a metric of the form \cite{Xie:2021bur}
\begin{align}\label{EQ:GenMetric}
ds^2 = & ~g_{tt}(r,\theta)dt^2 + g_{rr}(r,\theta)dr^2 + g_{\theta\theta}(r,\theta)d\theta^2 \nonumber\\&~
+ g_{\phi\phi}(r,\theta)d\phi^2 + 2 g_{t\phi}(r,\theta)dtd\phi.
\end{align}
In turn, this metric can be expressed in terms of an orthonormal tetrad $\{t_\alpha$, $r_\alpha$, $\theta_\alpha$, $\phi_\alpha\}$ in the following way 
\begin{align}
    g_{\alpha\beta} = - t_\alpha t_\beta + r_\alpha r_\beta +  \theta_\alpha \theta_\beta +  \phi_\alpha \phi_\beta,
\end{align}
where each of the vectors is spacelike except for $t_\alpha$, which is timelike.  While there is 
no unique choice of such an orthonormal tetrad, a convenient one for a metric of this form is 
\begin{align}\label{EQ:OrthoTetrad}
t_\alpha\partial^\alpha~=&~\sqrt{-g_{tt}}\partial^t-\frac{{g_{t\phi }}}{{\sqrt{-g_{tt}}}}\partial^\phi,\nonumber\\
r_\alpha\partial^\alpha~=&~\sqrt{g_{rr}}\partial^r,\nonumber\\
\theta_\alpha\partial^\alpha~=&~\sqrt{g_{\theta\theta}}\partial^\theta,\nonumber\\
\phi_\alpha\partial_\alpha~=&~-\frac{\sqrt{g_{t\phi }^2-g_{tt} g_{\phi \phi} }}{\sqrt{-g_{tt}}}\partial^\phi.
\end{align}

A complex null tetrad $\{l_\alpha$, $n_\alpha$, $m_\alpha$, $\bar{m}_\alpha\}$  of the Newman-Penrose formalism can then be constructed from the orthonormal one via \cite{Stephani_2003}
\begin{align}\label{EQ:NullTetrad}
l^\alpha=&~\frac{1}{\sqrt{2}} \left( t^\alpha + \phi^\alpha\right),\nonumber\\
n^\alpha=&~\frac{1}{\sqrt{2}} \left( t^\alpha - \phi^\alpha\right),\nonumber\\
m^\alpha=&~\frac{1}{\sqrt{2}} \left( r^\alpha + i ~\theta^\alpha\right),\nonumber\\
\bar{m}^\alpha=&~\frac{1}{\sqrt{2}} \left( r^\alpha - i ~\theta^\alpha\right).
\end{align}
Here, $l^\alpha$ and $n^\alpha$ are real and $m^\alpha$ and $\bar{m}^\alpha$ are a complex conjugate pair. All of the scalar products vanish except $l^\alpha n_\alpha = -1$ and $m^\alpha\bar{m}_\alpha = 1$ \footnote{this holds for the $(-,+,+,+)$ metric signature used in this paper, but the signs flip for the  $(+,-,-,-)$ signature}. 
This complex tetrad can be used to construct the metric via
\begin{align}
g_{\alpha\beta} = 2 m_{(\alpha} \bar{m}_{\beta)}- 2 l_{(\alpha} n_{\beta)} 
\end{align}
and contracted with the Weyl tensor $C_{\alpha\beta\gamma\delta}$ to compute the Weyl scalars 
\begin{align}
\Psi_0 =&~C_{\alpha\beta\gamma\delta}l^{\alpha}m^{\beta}l^{\gamma}m^{\delta},\nonumber\\
\Psi_1 =&~C_{\alpha\beta\gamma\delta}l^{\alpha}n^{\beta}l^{\gamma}m^{\delta},\nonumber\\
\Psi_2 =&~C_{\alpha\beta\gamma\delta}l^{\alpha}m^{\beta}\bar{m}^{\gamma}n^{\delta},\nonumber\\
\Psi_3 =&~C_{\alpha\beta\gamma\delta}l^{\alpha}n^{\beta}\bar{m}^{\gamma}n^{\delta},\nonumber\\
\Psi_4 =&~C_{\alpha\beta\gamma\delta}n^{\alpha}\bar{m}^{\beta}n^{\gamma}\bar{m}^{\delta}.
\end{align}
The complex null tetrad and the Weyl scalars are frame dependent quantities: there is not a unique choice of tetrad nor subsequently Weyl scalars for a given metric. An overview of Lorentz transformations in the Newman-Penrose formalism can be found in Appendix~\ref{SEC:LorentzTransformations}. Generally, $\Psi_0$ and $\Psi_4$ are associated with ingoing and outgoing transverse gravitational radiation, $\Psi_1$ and $\Psi_3$ are associated with ingoing and outgoing longitudinal radiation and $\Psi_2$ is associated with a Coulomb field~\cite{Szekeres_1965}.

With the Weyl scalar in hand, the Petrov type of a spacetime can be determined by the number of distinct roots $B$ of
\begin{align}\label{EQ:petrov}
\Psi_0 + 4B\Psi_1 + 6B^2\Psi_2 + 4B^3\Psi_3 + B^4\Psi_4 = 0,
\end{align}
computed in a Lorentz frame where $\Psi_4 \ne 0$, which, as it turns out, amounts to finding the reference frames in which $\Psi_0=0$. Such a frame  can also be found by preforming a Lorentz transformation to rotate the vector $l^\alpha$ to a new vector 
\begin{align}
k^\alpha = l^{\alpha} + \bar{B} m^\alpha + B\bar{m}^\alpha + B \bar{B} n^\alpha\,,
\end{align}
that satisfies
\begin{align}
    k^\alpha k^\beta k_{[\mu}C_{\nu]\alpha\beta[\rho}k_{\sigma]} = 0\,.
    \label{eq:k-PND}
\end{align}
We call this new vector $k^\alpha$ a PND.

In general, a spacetime has four PNDs corresponding to four distinct roots $B$ of the quartic Eq.~(\ref{EQ:petrov}) or alternatively, four roots of Eq.~\eqref{eq:k-PND}. When the four roots are distinct, the spacetime is said to be Petrov type I. However, it is possible that two or more of the roots coincide for a given spacetime. Such spacetimes are called algebraically special. In particular, a Petrov type D spacetime is an algebraically special spacetime with two doubly degenerate PNDs, i.e., there are only two distinct roots $B$ of Eq.~(\ref{EQ:petrov}). With the choice of frame given by Eqs. (\ref{EQ:OrthoTetrad}) and (\ref{EQ:NullTetrad}), $\Psi_1$ and $\Psi_3$ vanish in general, so the quartic Eq.~(\ref{EQ:petrov}) becomes a quadratic equation for $B^2$. When the discriminant  $9\Psi_2^2-\Psi_0\Psi_4$ vanishes, the spacetime is Petrov type D. Otherwise, there are four distinct roots and the spacetime is Petrov type I. This statement applies to any metric that can be written in the circular form of Eq. (\ref{EQ:GenMetric}).

While the tetrad and the Weyl scalars are frame dependent quantities, 
the PNDs are not and the Petrov type is an invariant way to classify spacetimes. The BHs of GR are Petrov type D and the Kerr metric possess the two doubly degenerate PNDs 
\begin{align}
k_{1,\mathrm{GR}}^\alpha\partial_\alpha~=&~\frac{r^2+a^2}{\Delta}\partial_t
+\partial_r
+\frac{a}{\Delta}\partial_\phi,  \nonumber\\
k_{2,\mathrm{GR}}^\alpha\partial_\alpha~=&~\frac{r^2+a^2}{\Delta}\partial_t
-\partial_r
+\frac{a}{\Delta}\partial_\phi.
\label{eq:k-GR}
\end{align}

The choice of frame given by Eqs.~(\ref{EQ:OrthoTetrad}) and (\ref{EQ:NullTetrad}) is convenient for determining the Petrov type and PNDs of a spacetime because only $\Psi_1$ and $\Psi_3$ vanish, and because it is simple to read off from a metric of the form given in Eq.~\eqref{EQ:GenMetric}. However, for a Petrov type D spacetime, it is possible to choose a frame such that $\Psi_0$, $\Psi_1$, $\Psi_3$, and $\Psi_4$ all vanish.
When working with BHs in GR, it is conventional to pick this frame because many equations simplify greatly. 

Let us then review how to transform the tetrad and Weyl scalars presented above to the preferred one, i.e.~that in which $\Psi_0=\Psi_1=\Psi_3=\Psi_4=0$, if the spacetime happens to be Petrov type D. Recall that such a spacetime has two doubly degenerate roots $B_1$ and $B_2$ of Eq. (\ref{EQ:petrov}). A class II transformation with parameter $B_1$, as specified in Appendix \ref{SEC:LorentzTransformations},  will set $\Psi_0 = \Psi_1 =0$. A subsequent transformation of class I with parameter $\bar{A} = (B_2-B_1)^{-1}$ will then set  $\Psi_3 = \Psi_4 =0$ ~\cite{Chandrasekhar_1992}. When, additionally, the remaining degrees of freedom, corresponding to a transformation of class III, are fixed so that the spin coefficient $\epsilon$ vanishes, the tetrad is now said to be the Kinnersley tetrad. Such a transformation rescales $l^{\alpha}$ and $n^{\alpha}$ and rotates $m^\alpha$ and $\bar{m}^\alpha$ without affecting any of the Weyl scalars. The Kinnersley tetrad for the Kerr BH is
\begin{align}\label{EQ:GRtetrad}
l^\alpha_\mathrm{GR}\partial_\alpha~=&~\frac{r^2+a^2}{\Delta}\partial_t
+\partial_r
+\frac{a}{\Delta}\partial_\phi,  \nonumber\\
n^\alpha_\mathrm{GR}\partial_\alpha~=&~\frac{r^2+a^2}{2\rho^2}\partial_t
-\frac{\Delta}{2\rho^2}\partial_r
+\frac{a}{2\rho^2}\partial_\phi, \nonumber\\
m^\alpha_\mathrm{GR}\partial_\alpha=&~\frac{1}{\sqrt{2}(r + ia\cos\theta)}\left(ia\sin\theta\partial_t + \partial_\theta + i\csc\theta\partial_\phi \right),
\end{align}
and the corresponding, nonvanishing  Weyl scalar is 
 \begin{align}\label{EQ:GRPsi2}
    \Psi_{2,\mathrm{GR}} = -\frac{M}{(r-ia\cos\theta)^3}.
\end{align}
It is the case that $l^\alpha$ and $n^\alpha$ are aligned with the PNDs in the Kinnersley frame, as shown above.

For a type I spacetime, which produces four roots $B_1$, $B_2$, $B_3$, and $B_4$ of Eq. (\ref{EQ:petrov}), it is not possible to pick a Kinnersley tetrad for which $\Psi_0$, $\Psi_1$, $\Psi_3$, and $\Psi_4$ vanish simultaneously and both $l^\alpha$ and $n^\alpha$ are aligned with PNDs. The best we can do is set $\Psi_0 = 0$ with a class II transformation with parameter $B_1$ and then set $\Psi_4=0$ with a class I transformation with parameter $\bar{A} = (B_2-B_1)^{-1}$, $\bar{A} = (B_3-B_1)^{-1}$, or $\bar{A} = (B_4-B_1)^{-1}$. In such a frame $l^\alpha$ is aligned with a PND but $n^\alpha$ is not. A class III rotation will now alter both the tetrad and the Weyl scalars. We will call the tetrad in such a frame ``Kinnersley-like.''
\subsection{Dynamical Chern-Simons Gravity}\label{SEC:dCSPNDs}
\begin{figure}
\centering
\includegraphics[width=8.5cm]{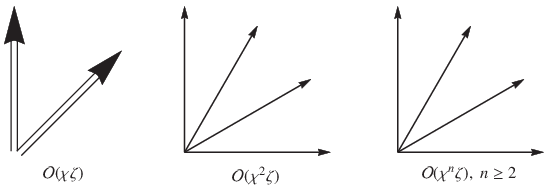}
\caption{\label{FIG:PNDs}Symbolic depictions of the PNDs of dCS gravity at different orders of spin in the metric. The dCS solution for a slowly rotating BH is Petrov type D at $\mathcal{O}(\chi\zeta)$ and therefore possesses two doubly degenerate PNDs. At $\mathcal{O}(\chi\zeta^2)$ the spacetime becomes Petrov type I and the two PNDs split into four. The spacetime will remain type I as higher orders in spin are included in the metric.}
\end{figure}

Let us now apply the above framework to spinning BHs in dCS gravity. The leading-order-in-$\zeta$, leading-order-in-$\chi$ dCS corrections to the PNDs are \begin{align}
k_{1,\mathrm{CS}}^\alpha\partial_{\alpha}=&~k_{1,\mathrm{GR}}^\alpha\partial_{\alpha} + \delta_\mathrm{CS}\partial_\phi +\mathcal{O}(\zeta\chi,\sqrt{\zeta}\chi^2), \nonumber\\
k_{2,\mathrm{CS}}^\alpha\partial_{\alpha}=&~k_{1,\mathrm{GR}}^\alpha\partial_{\alpha}- \delta_\mathrm{CS}\partial_\phi +\mathcal{O}(\zeta\chi,\sqrt{\zeta}\chi^2), \nonumber\\
k_{3,\mathrm{CS}}^\alpha\partial_{\alpha}=&~k_{2,\mathrm{GR}}^\alpha\partial_{\alpha}+ \delta_\mathrm{CS}\partial_\phi +\mathcal{O}(\zeta\chi,\sqrt{\zeta}\chi^2), \nonumber\\
k_{4,\mathrm{CS}}^\alpha\partial_{\alpha}=&~k_{2,\mathrm{GR}}^\alpha\partial_{\alpha}- \delta_\mathrm{CS}\partial_\phi +\mathcal{O}(\zeta\chi,\sqrt{\zeta}\chi^2),
\label{eq:dCSPND}
\end{align}
where
\begin{align}
 \delta_\mathrm{CS} = \frac{\sqrt{1407}}{112}\frac{\chi}{M}\sqrt{\zeta}\frac{M^2}{r^2}\Bigg(1 &+ \frac{2840}{603}\frac{M}{r} + \frac{64660}{4221}\frac{M^2}{r^2} \nonumber\\&
 + \frac{1740}{67}\frac{M^3}{r^3} + \frac{1980}{67}\frac{M^4}{r^4} \Bigg)^\frac{1}{2}.
\end{align}
We have checked that these $\mathcal{O}(\chi\sqrt{\zeta})$ corrections arise from contributions at $\mathcal{O}(\chi^2{\zeta})$ in the metric. 
The leading-order-in-coupling, first-order-in-spin metric therefore does not produce a dCS correction to the PNDs of GR and the spacetime  remains Petrov type D at that order, as reported by \cite{Sopuerta_2011}. At second order in spin in the dCS metric, the two PNDs of GR split into four, as shown above, making the spacetime Petrov type I and confirming the results of \cite{Yagi_2012}. The spacetime will remain Petrov type I as higher orders in spin are added to the metric. A symbolic depiction of the PNDs of dCS gravity at each order of spin in the metric is included in Fig. \ref{FIG:PNDs} .

The PNDs, and the roots $B$ we calculated to derive them, allow us to construct a Kinnersley-like complex tetrad.
As in the PND case, the metric at $\mathcal{O}(\chi^2{\zeta})$ sources $\mathcal{O}(\chi\sqrt{\zeta})$ corrections to this tetrad, namely 
\begin{align}\label{EQ:dCStetrad}
l^\alpha_\mathrm{CS}\partial_\alpha~=&~l^\alpha_\mathrm{GR}\partial_\alpha+\delta_\mathrm{CS}\partial_\phi + \mathcal{O}(\zeta\chi,\sqrt{\zeta}\chi^2),\nonumber\\
n^\alpha_\mathrm{CS}\partial_\alpha~=&~n^\alpha_\mathrm{GR}\partial_\alpha  +\frac{1}{2}f\delta_\mathrm{CS}\partial_\phi + \mathcal{O}(\zeta\chi,\sqrt{\zeta}\chi^2),\nonumber\\
m^\alpha_\mathrm{CS}\partial_\alpha=&~m^\alpha_\mathrm{GR}\partial_\alpha +i\frac{\sqrt{2}}{2}r\sin\theta~\delta_\mathrm{CS}\partial_t + \mathcal{O}(\zeta\chi,\sqrt{\zeta}\chi^2),
\end{align}
and the corresponding nonvanishing  Weyl scalars are
\begin{align}
\Psi_2^\mathrm{CS} =&~ \Psi_2^\mathrm{GR} + \cal{O}(\chi\zeta),  \nonumber\\
\Psi_1^\mathrm{CS}=&~ -\frac{2}{f}\Psi_3^\mathrm{CS} = -\frac{i3\sqrt{2}}{2}\frac{M^2}{r^2}\frac{\sin\theta}{M}\delta_\mathrm{CS} + \mathcal{O}(\chi^2\sqrt\zeta) \,,
\end{align}    
where $f= 1 -2M/r$. In general, the $\mathcal{O}(\chi^n\sqrt{\zeta})$ corrections to the PNDs, the tetrad and the Weyl scalars in this frame are sourced by terms of $\mathcal{O}(\chi^{n+1}\zeta)$ in the metric. Of course, the PNDs, the tetrad and the Weyl scalars do not just contain terms proportional to $\sqrt{\zeta}$ but also terms linear in $\zeta$ that must also be calculated. Therefore, because the metric is known to $\mathcal{O}(\chi^{5}\zeta)$, we are able to compute corrections to the PNDs, the tetrad and the Weyl scalars to  $\mathcal{O}(\chi^4\zeta)$. The corrections to $\mathcal{O}(\chi^2\zeta)$ are presented in Appendix \ref{SEC:Results} and the complete corrections are collected in a \textit{Mathematica} notebook that is provided in the Supplemental Material.

The above results allow for several conclusions. First, recall that the spacetime is not Petrov type D, so $n^{\alpha}_\mathrm{CS}$ is not aligned with a PND, and $\Psi_0^\mathrm{CS}, \Psi_1^\mathrm{CS}, \Psi_3^\mathrm{CS}, \Psi_4^\mathrm{CS}$ cannot all vanish simultaneously. While $\Psi_1^\mathrm{CS}$ and $\Psi_3^\mathrm{CS}$, which are associated with longitudinal gravitational waves, are nonzero, they are suppressed at future null infinity relative to the monopole term $\Psi_2^\mathrm{CS}$. This is why the only two gravitational wave polarization modes that survive at future null infinity are the two transverse-traceless ones, as in GR, consistent with the calculation of~\cite{Wagle:2019mdq}.

Second, note that there is not a unique choice of Kinnersley-like frame where $\Psi_0 = \Psi_4 =0$ and $l^\alpha_\mathrm{CS}$ is a PND of dCS gravity. Instead, what is shown above is one of 12 such frames, corresponding to four choices of class II transformation and subsequently three choices of class I transformation,  where this is the case. The remaining degrees of freedom, corresponding to a transformation of class III, were specified so that the GR part of the tetrad and the Weyl scalars are identical to
those
given in Eqs.~\eqref{EQ:GRPsi2} and~\eqref{EQ:GRtetrad}.

Third, note that the leading order correction to the quantities presented in this section are proportional to  $\sqrt{\zeta}$ while the leading order corrections to the metric are proportional to $\zeta$. This is to be expected because, for a perturbed type D spacetime, the leading order correction to the PNDs will be proportional to the square root of the perturbation parameter, which in our case is $\zeta$ \cite{Araneda_2015}. It is not possible to perform a transformation that will remove the $\sqrt\zeta$ proportional term of the Weyl scalars while preserving the GR parts and keeping $\Psi_0=\Psi_4=0$.

Finally, because both the spin parameter $\chi$ and coupling parameter $\zeta$ must be much smaller than unity (i.e.~we are here carrying out perturbative expansions in small spin and small coupling), it is not possible for a special case to occur in which terms of different orders in the expansion cancel each other. In particular, it is not possible for higher order terms not considered here to change the results presented in this section or the following ones.

\subsection{Scalar Gauss-Bonnet Gravity}\label{SEC:sGBPNDs}
\begin{figure}
\centering
\includegraphics[width=8.5cm]{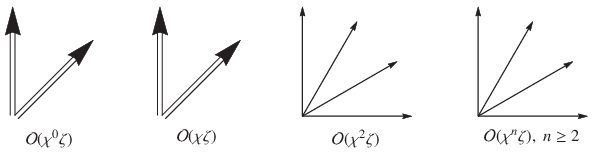}
\caption{\label{FIG:PNDsGB}Symbolic depictions of the PNDs of sGB gravity at different orders of spin in the metric. The sGB solution for a slowly rotating BH is Petrov type D at $\mathcal{O}(\chi^0\zeta)$ and $\mathcal{O}(\chi\zeta)$ therefore possesses 2 doubly degenerate PNDs at these orders. At $\mathcal{O}(\chi\zeta^2)$ the spacetime becomes Petrov type I and the two PNDs split into four. The spacetime will remain type I as higher orders in spin are included in the metric.}
\end{figure}

In the same way, we can compute the PNDs, tetrad and Weyl scalars of sGB gravity. The leading-order-in-$\zeta$, leading-order-in-$\chi$ sGB corrections to the PNDs are 
\begin{align}
k_{1,\mathrm{GB}}^\alpha\partial_{\alpha}=&~k_{1,\mathrm{GR}}^\alpha\partial_{\alpha} + \delta_\mathrm{GB}\partial_\theta + \mathcal{O}(\zeta,\sqrt{\zeta}\chi^2), \nonumber\\
k_{2,\mathrm{GB}}^\alpha\partial_{\alpha}=&~k_{1,\mathrm{GR}}^\alpha\partial_{\alpha}- \delta_\mathrm{GB}\partial_\theta + \mathcal{O}(\zeta,\sqrt{\zeta}\chi^2), \nonumber\\
k_{3,\mathrm{GB}}^\alpha\partial_{\alpha}=&~k_{2,\mathrm{GR}}^\alpha\partial_{\alpha}+ \delta_\mathrm{GB}\partial_\theta + \mathcal{O}(\zeta,\sqrt{\zeta}\chi^2), \nonumber\\
k_{4,\mathrm{GB}}^\alpha\partial_{\alpha}=&~k_{2,\mathrm{GR}}^\alpha\partial_{\alpha}- \delta_\mathrm{GB}\partial_\theta + \mathcal{O}(\zeta,\sqrt{\zeta}\chi^2),
\end{align}

where
\begin{align}
\delta_\mathrm{GB} =&~ \frac{\sqrt{468615}}{525} \frac{\chi}{M}\sqrt{\zeta}\frac{M^2}{r^2}\sin\theta\left(1+\frac{21440}{4463}\frac{M}{r}\right.\nonumber\\&~\left.+\frac{508960}{31241}\frac{M^2}{r^2}+\frac{135300}{4463}\frac{M^3}{r^3}+\frac{167600}{4463}\frac{M^4}{r^4}\right)^{\frac{1}{2}}.
\end{align}
As was the case for dCS gravity, these $\mathcal{O}(\chi\sqrt{\zeta})$ corrections contain contributions from the $\mathcal{O}(\chi^2{\zeta})$ term of the metric. 
Neither the nonspinning metric nor the first-order-in-spin metric in the small coupling regime of sGB gravity produces corrections to the PNDs of GR and the spacetime  remains Petrov type D at these orders. At second order in spin in the sGB metric, the two PNDs of GR split into four as shown above, making the spacetime Petrov type I. This Petrov classification agrees with what is given in  \cite{Ayzenberg_2014}. 
The spacetime will remain Petrov type I as higher orders in spin are added to the metric. A symbolic depiction of the PNDs of sGB gravity at each order of spin in the metric is included in Fig. \ref{FIG:PNDsGB}. 

Once again, with the PNDs in hand, we can compute the complex tetrad in a Kinnersley-like frame.  The metric at $\mathcal{O}(\chi^2\zeta)$ sources $\mathcal{O}(\chi\sqrt\zeta)$ corrections to the tetrad, leading to 
\begin{align}\label{EQ:sGBtetrad}
l^\alpha_\mathrm{GB}\partial_\alpha~=&~l^\alpha_\mathrm{GR}\partial_\alpha+\delta_\mathrm{GB}\partial_\theta + \mathcal{O}(\zeta,\sqrt{\zeta}\chi^2),\nonumber\\
n^\alpha_\mathrm{GB}\partial_\alpha~=&~n^\alpha_\mathrm{GR}\partial_\alpha - \frac{1}{2}f\delta_\mathrm{GB}\partial_\theta+ \mathcal{O}(\zeta,\sqrt{\zeta}\chi^2),\nonumber\\
m^\alpha_\mathrm{GB}\partial_\alpha=&~m^\alpha_\mathrm{GR}\partial_\alpha -\frac{\sqrt2}{2}\delta_{GB}\left(rf\partial_r-i\sqrt{f}\partial_{\theta}\right.\nonumber\\&+\frac{1}{\sin\theta}\sqrt{f}\partial_{\phi}\Big)+ \mathcal{O}(\zeta,\sqrt{\zeta}\chi^2),
\end{align}
and the nonvanishing  Weyl scalars
\begin{align}
\Psi_2^\mathrm{GB} =&~ \Psi_2^\mathrm{GR}+ \mathcal{O}(\zeta),  \nonumber\\
\Psi_1^\mathrm{GB}=&~ -\frac{2}{f}\Psi_3^\mathrm{CS} = -\frac{3\sqrt{2}}{2}\frac{M^2}{r^2}\frac{1}{M}\delta_\mathrm{GB}+ \mathcal{O}(\chi^2\sqrt\zeta).
\end{align} 
As before, the $\mathcal{O}(\chi^n\sqrt{\zeta})$ corrections to the PNDs, the tetrad and the Weyl scalars in this frame are sourced by terms of $\mathcal{O}(\chi^{n+1}\zeta)$ in the metric. With a metric to $\mathcal{O}(\chi^{5}\zeta)$, we are able to compute corrections to the PNDs, the tetrad and the Weyl scalars to $\mathcal{O}(\chi^4\zeta)$. Again, the corrections to $\mathcal{O}(\chi^2\zeta)$ are presented in Appendix \ref{SEC:Results} and the complete corrections are collected in a \textit{Mathematica} notebook that is provided in the Supplemental Material.

As in the dCS case, we can extract similar conclusions about the mathematical properties of the sGB solutions. We see that $n_\alpha^\mathrm{GB}$ is no longer aligned with a PND and $\Psi_2$ can no longer be the only nonvanishing  Weyl scalar. As before, the Weyl scalars associated with radiation are suppressed at infinity relative to the monopole term, once more yielding results consistent with~\cite{Wagle:2019mdq}.  Finally, what is shown above is just one choice of frame where the GR part of the tetrad and Weyl scalars matches what is given in  Eqs.~\eqref{EQ:GRPsi2} and~\eqref{EQ:GRtetrad}.

\section{\label{SEC:Conserved Quantities}Conserved Quantities}
Let us now turn our attention to Killing tensors. A  Killing tensor $K_{\mu_1..\mu_N}$ of rank $N$ is a completely symmetric tensor that obeys the Killing equation 
\begin{equation}
    \nabla_{(\nu}K_{\mu_1..\mu_N)} = 0. 
\end{equation}
Killing tensors are of particular interest because the scalar quantity $K_{\mu_1..\mu_N}u^{\mu_1}...u^{\mu_N}$ is conserved along the geodesic $u^\mu$, which by definition satisfies $u^\beta\nabla_\beta u^\mu = 0.$

For a test particle moving along a geodesic in a stationary, axially symmetric spacetime, energy $\tilde{E}$, the component of angular momentum along the axis of symmetry  $\tilde{L}$, and rest mass are conserved. These conserved quantities are associated with the Killing vectors $t^\alpha\partial_\alpha = \partial_t$ and $\phi^\alpha\partial_\alpha = \partial_\phi$ and the metric $g_{\mu\nu}$, which is a Killing tensor by virtue of metric compatibility. The Kerr metric of GR also posseses a fourth conserved quantity known as the Carter constant. The Carter constant 
\begin{equation}
    \mathcal{Q} = \xi_{\alpha\beta}^\mathrm{GR}u^\alpha u^\beta -(\tilde{L} - a \tilde{E})^2  
\end{equation}
is generated by the rank-2 Killing tensor 
\begin{equation}
    \xi_{\alpha\beta}^\mathrm{GR} = \Delta \,  k^{1,\mathrm{GR}}_{(\alpha}k^{2,\mathrm{GR}}_{\beta)} + r^2g^{\rm GR}_{\alpha\beta}\,,
\end{equation}
where $k_{1,2,\rm GR}^{\alpha}$ were already given in Eq.~\eqref{eq:k-GR}
and $\Delta$ is the metric function given in Eq.~\eqref{EQ:Kerrmetricfunctions}.

The existence of such a rank-2 Killing tensor is guaranteed for Petrov type D spacetimes as the Kerr metric of GR, but both modified theories we are considering here are type I beginning at second order in spin. The existence of a Killing tensor, and thus of a fourth conserved quantity, is neither guaranteed nor ruled out for such spacetimes.  

Without an additional Killing tensor to generate a fourth symmetry, the equations of motion cannot be put into quadrature form and the geodesic motion is considered to be Liouville chaotic~\cite{Schuster_2006}. Chaotic features of geodesics on BH backgrounds may be encoded in gravitational wave signals from extreme mass-ratio inspirals observable by future space-based detectors, such as the Laser Interferometer Space Antenna (LISA) \cite{Kaiser:2020tlg}. Such observations could allow us to place constraints on deviations from GR or to perform null tests. This motivates the following investigation into the existence of a Killing tensor that would generate a fourth constant of the motion for each modified theory considered here.

\subsection{Dynamical Chern-Simons Gravity}
As the Kerr metric of GR is, the BH of dCS gravity is stationary and axially symmetric, and therefore admits the Killing vectors $t^\alpha$ and $\phi^\alpha$ in addition to the metric which is always a Killing tensor. The spacetime is also Petrov type D through $\mathcal{O}(\chi\zeta)$ so we expect it to possess an extension of the rank-2 Killing tensor of GR through that order as well. That Killing tensor is \cite{Sopuerta_2011}
\begin{align}
\label{EQ:dCSRank2}
 \xi_{\alpha\beta}^\mathrm{CS} = \Delta \,  \bar{k}^{1,\mathrm{CS}}_{(\alpha}\bar{k}^{2,\mathrm{CS}}_{\beta)} + r^2g_{\alpha\beta}^\mathrm{CS},
\end{align}
where
\begin{align} \label{EQ:dCSkbar}
    \bar{k}^{1,\mathrm{CS}}_{\alpha} = k^{1,\mathrm{GR}}_{\alpha} + \delta\bar k_\alpha^\mathrm{CS},\nonumber\\
    \bar{k}^{2,\mathrm{CS}}_{\alpha} = k^{2,\mathrm{GR}}_{\alpha} + \delta \bar k_\alpha^\mathrm{CS},
\end{align}
and 
\begin{align}
\delta \bar k^\alpha_\mathrm{CS} \partial_\alpha = -\frac{5}{8}\chi\zeta\frac{M^6}{r^6}\frac{1}{Mf} \left(1 + \frac{12}{7}\frac{M}{r}+\frac{27}{12}\frac{M^2}{r^2}\right).
\end{align}
The $\bar{k}_{1,2,\rm CS}^\alpha$ in Eq.~\eqref{EQ:dCSkbar} are \textit{not} the same as the $k_{1,2,\rm CS}^\alpha$ given  Eq.~\eqref{eq:dCSPND}, and in fact are \textit{not} PNDs of the spacetime.
One may then wonder whether it is possible to write the rank-2 Killing tensor given in Eq.~\eqref{EQ:dCSRank2} in terms of the dCS PNDs presented in Eq.~\eqref{eq:dCSPND} via an ansatz such as
\begin{align}
\label{EQ:dCS-ansatz}
 \xi_{\alpha\beta}^\mathrm{CS} =&~
 +F_1 k^{1,\mathrm{CS}}_{(\alpha} k^{1,\mathrm{CS}}_{\beta)}
 +F_2 k^{1,\mathrm{CS}}_{(\alpha} k^{2,\mathrm{CS}}_{\beta)}
 +F_3 k^{1,\mathrm{CS}}_{(\alpha} k^{3,\mathrm{CS}}_{\beta)}\nonumber\\&~ 
 +F_4 k^{1,\mathrm{CS}}_{(\alpha} k^{4,\mathrm{CS}}_{\beta)}
 +F_5 k^{2,\mathrm{CS}}_{(\alpha} k^{2,\mathrm{CS}}_{\beta)}  
 +F_6 k^{2,\mathrm{CS}}_{(\alpha} k^{3,\mathrm{CS}}_{\beta)}\nonumber\\&~ 
 +F_7 k^{2,\mathrm{CS}}_{(\alpha} k^{4,\mathrm{CS}}_{\beta)}
 +F_8 k^{3,\mathrm{CS}}_{(\alpha} k^{3,\mathrm{CS}}_{\beta)}
 +F_9 k^{3,\mathrm{CS}}_{(\alpha} k^{4,\mathrm{CS}}_{\beta)}\nonumber\\&~
 +F_{10} k^{4,\mathrm{CS}}_{(\alpha} k^{4,\mathrm{CS}}_{\beta)}
 +F_{11} g_{\alpha\beta}^\mathrm{CS},
\end{align}
where $F_i = F_i(r,\theta)$. However, there is no choice of functions $F_i(r,\theta)$ that will produce Eq. \eqref{EQ:dCSRank2}. Moreover, Ref.~\cite{Yagi_2012} showed that there is no analogous extension to Eq.~\eqref{EQ:dCSkbar} at $\mathcal{O}(\chi^2\zeta)$, precluding the existence of a rank-2 Killing tensor and a Carter-like constant at this order, and therefore, at any other higher order in spin as well.

With the knowledge that the dCS BH does not possess a rank-2 Killing tensor nor a Carter-like constant, C\'ardenas-Avenda\~no \textit{et al.}~\cite{Cardenas_Avendano_2018} 
explored the existence of a fourth symmetry numerically. They did so by searching for chaos in geodesic motion outside such BHs in the slow-rotation approximation. While chaos was found, it was shown to diminish as higher orders in spin were added to the BH background used in the simulations. A similar phenomenon was observed in the slow rotation approximation of GR, where it is known that the full solution is nonchaotic.  The authors therefore conjectured that the exact solution in dCS gravity would also be nonchaotic, indicating the existence of a fourth constant of motion. 

Such a constant must be generated by a Killing tensor but, with the existence of a rank-2 Killing tensor already ruled out~\cite{Yagi_2012}, 
we must look for higher-rank Killing tensors. 
The logical place to start the search is with rank-3 Killing tensors. We 
take the most general ansatz, written in the form
\begin{align}\label{EQ:Rank3Ansatz}
K_{\alpha\beta\gamma} = &~
K_{\alpha\beta\gamma}^{(0,0)} +\chi' K_{\alpha\beta\gamma}^{(1,0)} +\chi'^2K_{\alpha\beta\gamma}^{(2,0)}\nonumber\\
&~+\zeta'\left(\chi' K_{\alpha\beta\gamma}^{(1,1)} +\chi'^2K_{\alpha\beta\gamma}^{(2,1)}\right),
\end{align}
where $\chi'$ and $\zeta'$ are bookkeeping parameters that label the orders of the slow-rotation and small-coupling approximations respectively, and the $K_{\alpha\beta\gamma}^{(m,n)}$ are completely symmetric tensor fields that depend on only $r$ and $\theta$ and are  proportional to $\chi^m\zeta^n$. In this way, the tensors $K_{\alpha\beta\gamma}^{(0,0)}$, $K_{\alpha\beta\gamma}^{(1,0)}$ and $K_{\alpha\beta\gamma}^{(2,0)}$ are GR contributions, while $K_{\alpha\beta\gamma}^{(1,1)}$ and $K_{\alpha\beta\gamma}^{(2,1)}$ would be dCS corrections.

We substitute the above ansatz into the Killing equation and solve order by order to determine the most general rank-3 Killing tensor of the spacetime.
A linear combination of symmetrized exterior products of Killing tensors is always a Killing tensor itself. Indeed, after solving the Killing equation order by order, we find that the solution through $\mathcal{O}(\chi\zeta)$, but not including the $\mathcal{O}(\chi^2\zeta)$ terms, is
\begin{align}\label{EQ:genrank3}
      &K_{\alpha\beta\gamma}^{(0,0)} +  K_{\alpha\beta\gamma}^{(1,0)} +  K_{\alpha\beta\gamma}^{(2,0)} +  K_{\alpha\beta\gamma}^{(1,1)}=  \nonumber \\ &~   
        C_1 t_{\alpha}t_{\beta}t_{\gamma} 
    +   C_2 t_{(\alpha}t_{\beta}\phi_{\gamma)} 
    +   C_3  t_{(\alpha}\phi_{\beta}\phi_{\gamma)} \nonumber\\&~
    +   C_4 \phi_{(\alpha}\phi_{\beta}\phi_{\gamma)} 
    +   C_5 t_{(\alpha}g_{\alpha\beta)} 
    +   C_6 \phi_{(\alpha}g_{\alpha\beta)} \nonumber\\&~
    +   C_7 t_{(\alpha}\xi^\mathrm{CS}_{\alpha\beta)} 
    +   C_8 \phi_{(\alpha}\xi^\mathrm{CS}_{\alpha\beta)},
\end{align}
where the $C_i$ are arbitrary constants and each term is understood to be expanded about small spin and coupling to $\mathcal{O}(\chi\zeta)$. We see that this Killing tensor is nothing but a linear combination of symmetrized exterior products of the $\mathcal{O}(\chi\zeta)$ Killing vectors $t^\alpha$ and $\phi^\alpha$, and the $\mathcal{O}(\chi\zeta)$ Killing tensors $\xi^{CS}_{\alpha \beta}$ and $g_{\alpha \beta}$. These four quantities already generate four constants of the motion at $\mathcal{O}(\chi\zeta)$, and therefore, Eq.~\eqref{EQ:genrank3} does not generate a new independent constant at this order, but rather generates a constant that is a function of the other constants. 

Can we now solve the Killing equation at $\mathcal{O}(\chi^2\zeta)$ and extend the fourth constant of motion to that order? What this means is that we must use Eq.~\eqref{EQ:genrank3} through $\mathcal{O}(\chi\zeta)$ and $K_{\alpha\beta\gamma}^{(2,1)}$ to compose $K_{\alpha\beta\gamma}$ and require that $K_{(\alpha\beta\gamma;\delta)} = 0$. The terms through $\mathcal{O}(\chi\zeta)$ generate $\mathcal{O}(\chi^2\zeta)$ terms in the Killing equation, and these must be canceled by derivatives of $K_{\alpha\beta\gamma}^{(2,1)}$. Since the latter is a symmetric rank-3 tensor, we have $20$ functions of $(r,\theta)$ to determine from $35$ components of the Killing equations, which implies the system may then be overdetermined. In practice, we find that this is the case because, even though some of the $35$ components of the equation are trivially satisfied, some of the $20$ components of $K_{\alpha\beta\gamma}^{(2,1)}$ must be set to zero due to symmetry\footnote{The even and odd parts (under simultaneous time and azimuthal angle reflection) of a given conserved quantity must be conserved independently. This implies that the even and odd Killing tensors that generate these conserved quantities must also satisfy the Killing equations independently. Since at lower spin-order the most general rank-3 Killing tensor of dCS gravity is entirely odd, it must remain so at higher spin-order. In practice, this means that certain components of the correction to the Killing tensor, such as $K_{ttr}^{(2,1)}$, must be set to zero.}. In fact, we find that the system of Killing equations is \textit{inconsistent}; i.e.~a subset of the components of the Killing equations require a certain functional form for certain components of $K_{\alpha\beta\gamma}^{(2,1)}$, which is then not allowed by another subset of components of the Killing equations. We will not show this explicitly here because it is un-illuminating, but we have verified it by hand, in \textit{Maple} and in \textit{Mathematica}.  

We have generalized the above results to Killing tensors of higher rank. To do so, we began by generalizing the ansatz in Eq. (\ref{EQ:Rank3Ansatz}) to tensors of rank 4, 5 and 6. Through $\mathcal{O}(\chi\zeta)$ we again find that the most general solution is given by symmetrized exterior products of the two Killing vectors and the two Killing tensors at $\mathcal{O}(\chi\zeta)$. When we then attempt to solve the Killing equations at $\mathcal{O}(\chi^2\zeta)$, we again find that the system is inconsistent in  the sense described above. With this, we have established that no Killing tensor of rank 2, 3, 4, 5, or 6 exists in dCS gravity at $\mathcal{O}(\chi^2\zeta)$, precluding the existence of one at higher orders in spin as well. We therefore conjecture, though we cannot prove, that no Killing tensor of any rank exists in dCS gravity, and thus, that there does not exist a fourth constant of the motion and the geodesic motion should be Liouville chaotic. 

\subsection{Scalar Gauss Bonnet Gravity}
The BH of sGB gravity is also Petrov type D through $\mathcal{O}(\chi\zeta)$ and therefore we expect it to have an extension to the rank-2 Killing tensor of GR through this order. 
We search for that Killing tensor by taking an ansatz analogous to Eq.~\eqref{EQ:Rank3Ansatz}
\begin{align}\label{EQ:Rank2Ansatz}
K_{\alpha\beta} = &~
K_{\alpha\beta}^{(0,0)} +\chi' K_{\alpha\beta}^{(1,0)} +\chi'^2K_{\alpha\beta}^{(2,0)}\nonumber\\&~
+\zeta' K_{\alpha\beta}^{(0,1)}
+\zeta'\chi' K_{\alpha\beta}^{(1,1)},
\end{align}
where again $K_{\alpha\beta}^{(m,n)}$ are completely symmetric tensor fields that depend on only $r$ and $\theta$ and are  proportional to $\chi^m\zeta^n$. We substitute this ansatz into the Killing equation and find that the most general rank-2 Killing tensor of sGB gravity is
\begin{align}\label{EQ:sGBgenrank2}
      K_{\alpha\beta} = &~ 
        C_1 t_{\alpha}t_{\beta}
    +   C_2 t_{(\alpha}\phi_{\beta)} 
    +   C_3  \phi_{\alpha}\phi_{\beta}\nonumber\\ 
   &~ +   C_4 g_{\alpha\beta} 
    +   C_5 \xi^\mathrm{GB}_{\alpha\beta}, 
\end{align}
where the $C_i$ are arbitrary constants and each term is understood to be expanded about small spin and coupling to $\mathcal{O}(\chi\zeta)$.
The last term of Eq. \eqref{EQ:sGBgenrank2} constitutes an extension of $\xi^\mathrm{GR}_{\alpha\beta}$ and generates a correction to the Carter constant at $\mathcal{O}(\chi\zeta)$.  It is given by
\begin{equation}\label{EQ:sGBRank2}
 \xi_{\alpha\beta}^\mathrm{GB} = \Delta \bar{k}^{1,\mathrm{GB}}_{(\alpha}\bar{k}^{2,\mathrm{GB}}_{\beta)} + r^2g^\mathrm{GB}_{\alpha\beta},
\end{equation}
where
\begin{align}
    \bar{k}^{1,\mathrm{GB}}_{\alpha} = k^{1,\mathrm{GR}}_{\alpha} + \delta\bar k^{1,\mathrm{GB}}_\alpha,\nonumber\\
    \bar{k}^{2,\mathrm{GB}}_{\alpha} = k^{2,\mathrm{GR}}_{\alpha} + \delta\bar k^{2,\mathrm{GB}}_\alpha.\nonumber\\
\end{align}
The nonzero components of $\delta\bar k^{1,2\mathrm{GB}}_\alpha$ are 
\begin{widetext}
\begin{align}
   \delta\bar k^{1,\mathrm{GB}}_{t} =&~  \delta\bar k^{2,\mathrm{GB}}_{t} = 
   -\frac{1}{6}\zeta\frac{1}{f^2} \frac{M^3}{r^3}\left(1+26\frac{M}{r}+\frac{66}{5}\frac{M^2}{r^2}+\frac{96}{5}\frac{M^3}{r^3}-80\frac{M^2}{r^2}\right),\nonumber\\ \delta \bar k^{1,\mathrm{GB}}_{r} =&~  -\delta \bar k^{2,\mathrm{GB}}_{r} =-\frac{1}{2}\zeta\frac{1}{f} \frac{M^2}{r^2}\left(1+\frac{M}{r}+\frac{52}{3}\frac{M^2}{r^2}+2\frac{M^3}{r^3}+\frac{16}{5}\frac{M^2}{r^2}-\frac{368}{3}\frac{M^2}{r^2}\right),\nonumber\\
  \delta \bar k^{1,\mathrm{GB}}_{\phi} =&~  \delta \bar k^{2,\mathrm{GB}}_{\phi} = -\frac{13}{30}\chi\zeta\frac{1}{Mf^2}\frac{M^5}{r^5}\left(1+\frac{134}{13}\frac{M}{r}+\frac{74}{13}\frac{M^2}{r^2}+\frac{96}{13}\frac{M^3}{r^3}-\frac{592}{13}\frac{M^4}{r^4}\right). 
\end{align}
\end{widetext}
Again, note that the  $\bar{k}^{1,2\mathrm{GB}}_{\alpha}$ are not PNDs of sGB gravity, and it is not possible to write the Killing tensor given in Eq.~\eqref{EQ:sGBRank2} in terms of the sGB PNDs via an ansatz analogous to Eq. \eqref{EQ:dCS-ansatz}.

The question now is whether a Killing tensor exists at $\mathcal{O}(\chi^2\zeta)$. To investigate this question, we follow the same approach as in dCS gravity and consider a generic $K_{\alpha\beta}^{(2,1)}$ correction to Eq.~\eqref{EQ:Rank2Ansatz}. This correction must be determined by solving the Killing equation $K_{(\alpha \beta;\gamma)} = 0$ to $\mathcal{O}(\chi^2\zeta)$. As in the dCS case, however, we find that the system of partial differential equations produced by the Killing equation is inconsistent; i.e. a subset of the equations requires a certain functional form for $K_{\alpha\beta}^{(2,1)}$, which is incompatible with a different subset of the equations. This implies that a Killing tensor of rank 2 does not exist in sGB gravity, and therefore, there is no Carter-like constant associated with one. 
This does not necessarily imply that a higher-rank Killing tensor does not exist in sGB gravity. However, as we saw in dCS gravity, higher-rank tensors do not necessarily enable the system of Killing equations to be consistent, and thus, it is likely that a fourth conserved quantity does not exist at all in either dCS gravity or sGB gravity.  

\section{\label{SEC:Conclusion}Conclusion}
In this paper, we have computed the PNDs of dCS and sGB gravity to $\mathcal{O}(\chi^5\zeta)$, confirming that both spacetimes are Petrov type I. We have also computed the Weyl scalars and complex null tetrad for both theories to the same order in a frame where $l^\alpha$ is a PND and $\Psi_0 = \Psi_4 =0$, showing that not only is $\Psi_2 \neq 0$ but also $\Psi_1 \neq 0 \neq \Psi_3$. As a bonus while carrying out these calculations, we also described a method that can be used to compute the PNDs, complex null tetrad and Weyl scalars in a large class of BH spacetimes.

These results have important implications to the study of vacuum perturbations of dCS and sGB BHs. In contrast to the BHs of dCS and sGB gravity, BHs in GR are Petrov type D, and it is therefore possible to choose a Kinnersley tetrad such that both $l^\alpha$ and $n^\alpha$ are aligned with the PNDs of the spacetime and the only nonvanishing  Weyl scalar is $\Psi_2$.  These were important assumptions in the derivation of the Teukolsky equations for BH perturbations. Because spinning BHs in both dCS and sGB gravity are Petrov type I, these assumptions are not valid in these theories. The quantities presented in this paper, however, can be used to extend the framework to these non-Petrov type D spacetimes. 

The Weyl scalars encode information about gravitational radiation. Specifically, $\Psi_0$ and $\Psi_4$ are associated with ingoing and outgoing transverse gravitational radiation, $\Psi_1$ and $\Psi_3$ are associated with ingoing and outgoing longitudinal radiation and $\Psi_2$ is associated with a Coulomb field. Although the Weyl scalars $\Psi_{1}$ and $\Psi_{3}$ do not vanish in the quadratic gravity theories considered here, they fall off faster than $r^{-1}$. That is, the only relevant radiative degrees of freedom at future null infinity (where gravitational waves are measured) are the gravitational wave polarizations encoded in $\Psi_{4}$. This is consistent with the analysis developed in~\cite{Wagle:2019mdq}.

Quadratic gravity theories such as dCS and sGB often find motivation in the low-energy limit of string theory. It would therefore be interesting to consider the Petrov type of the corresponding higher-dimensional theories; see e.g. Ref.~\cite{Coley:2004jv} for a classification of the Weyl tensor.
Most analyses have focused on black hole solutions to GR in higher dimensions~\cite{Durkee:2010qu,Coley:2011dm,Godazgar:2011sn,Coley:2012gn,Dias:2013hn} or a subclass of solutions in quadratic gravity complementary to our study~\cite{Pravda:2016fue,Podolsky:2019gro}. 
We expect that the Petrov type may be conserved under dimensional reduction or a Kaluza-Klein compactification under certain conditions. However, a proof of this would require an analysis of the  phasespace of solutions, including rotating solutions, in quadratic gravity in $D$ dimensions. This is a possible avenue for future research.

We have also explored the existence of a fourth constant of the motion in both dCS and sGB gravity by searching for new independent Killing tensors. While the spinning BHs of dCS gravity do possess an independent rank-2 Killing tensor to $\mathcal{O}(\chi\zeta)$~\cite{Sopuerta_2011}, it had been previously shown that one does not exist at $\mathcal{O}(\chi^2\zeta)$~\cite{Yagi_2012}. We have extended this result to show that there is no independent Killing tensor up to and including rank-6 to $\mathcal{O}(\chi^2\zeta)$. 
Even though we cannot prove that this result continues to hold for Killing tensors of rank higher than 6, we deem this possibility likely. 
We therefore conjecture, excluding the unlikely possibility of the existence of a conserved quantity not generated by a Killing tensor, that the spinning BHs of dCS gravity do not possess a fourth constant of motion and geodesic motion on such backgrounds should be chaotic. 

This prediction is in contrast to the conjecture in \cite{Cardenas_Avendano_2018}. That study searched numerically for chaos in geodesic motion on a spinning BH background in dCS gravity. The authors showed that, while such geodesic motion is chaotic, as more terms are included in an expansion about small spin, the chaos shrinks. Because a similar phenomenon was observed for the small spin expansion of the Kerr metric, which does possess a fourth constant of the motion, it was argued that the dCS metric likely possesses  such a constant as well. Our results suggest that, while chaos in geodesic motion in dCS gravity might reduce as higher order spin terms are included, it will not converge to zero as is the case for GR. Therefore, there may have been a remnant of chaos in the geodesics studied in~\cite{Cardenas_Avendano_2018} but of perhaps too small a size to be resolved numerically. 

For sGB gravity, we have computed an independent, rank-2 Killing tensor to $O(\chi\zeta)$ but have found that there is no independent rank-2 Killing tensor at $O(\chi^2\zeta)$. 
As in the dCS case, this implies, though it does not prove, that an exact BH solution in sGB gravity may not possess fourth constant of motion either. It would therefore be interesting to repeat the search for chaos carried out for dCS gravity for either theory using an exact numerical metric. If chaos was found in such a study, it would definitively rule out the existence of a fourth constant of motion in the given theory. This, in turn, would prove that no Killing tensor of any rank exists for such metrics. 

\acknowledgements
We thank Alejandro C\'ardenas-Avenda\~no, Alexander Deich, and Leo C. Stein, for useful discussions and  Dimitry Ayzenberg for providing the sGB metric in Boyer-Lidquist coordinates.
%
N.Y.~acknowledges financial support through NSF Grants
No. PHY-1759615, No. PHY-1949838 and NASA ATP Grant No. 17-ATP17-0225, No.~NNX16AB98G and
No.~80NSSC17M0041.
H.W.~acknowledges financial support provided by NSF Grant No. OAC-2004879 and
support by the Royal Society Research Grant No.~RGF\textbackslash R1\textbackslash 180073.

\appendix
\section{Slow-Rotation, Small-Coupling BH Solutions in Boyer-Lindquist Coordinates}\label{SEC:dCSMetric}
\subsection{Dynamical Chern-Simons Gravity}
\newpage
Here we include the slow-rotation, small-coupling BH solution of dCS gravity in Boyer Lindquist coordinate
\begin{align}
    g_{\alpha\beta}^\mathrm{CS} = g_{\alpha\beta}^{\mathrm{GR}} + \delta g_{\alpha\beta}^\mathrm{CS}, 
\end{align}
where $g^\mathrm{GR}_{\alpha\beta}$ refers to the Kerr metric of GR, Eq.~\eqref{EQ:Kerrmetric},
expanded for $\chi \ll 1.$

The first-order-in-spin term of the dCS correction $\delta g_{\alpha\beta}^\mathrm{CS}$, presented below, was derived by \cite{Yunes_2009} and the second-order-in-spin term by \cite{Yagi_2012}. The remaining terms were computed by \cite{Maselli_2017} in Hartle-Thorne coordinates and transformed to Boyer-Lindquist coordinates by \cite{Cardenas_Avendano_2018}.  
\begin{widetext}
\begin{eqnarray}
\delta g_{tt}^\mathrm{CS}
=& ~~~-\zeta \chi^2 & \frac{M^3}{r^3} \Bigg[ \frac{5}{384} \frac{M^2}{r^2} \left( 1 + 100 \frac{M}{r} + 194\frac{M^2}{r^2}  \frac{2220}{7} \frac{M^3}{r^3} \right.  \left. - \frac{1512}{5} \frac{M^4}{r^4} \right)\nonumber\\
&&-\frac{201}{1792} \left( 1+\frac{M}{r} +\frac{4474}{4221} \frac{M^2}{r^2} -\frac{2060}{469} \frac{M^3}{r^3} +\frac{1500}{469} \frac{M^4}{r^4}  - \frac{2140}{201} \frac{M^5}{r^5} + \frac{9256}{201} \frac{M^6}{r^6} - \frac{5376}{67} \frac{M^7}{r^7}  \right) (3\cos^2 \theta -1)\Bigg] 
\nonumber\\
 &+\zeta \chi^4&  \frac{M^3}{r^3}  \Bigg[ \frac{1}{384} \frac{M^2}{r^2}  \left(1+\frac{2624}{35} \frac{M}{r}+\frac{492831}{3920} \frac{M^2}{r^2}+\frac{1771487}{1680} \frac{M^3}{r^3} +\frac{330775}{168} \frac{M^4}{r^4} +\frac{4430511}{980} \frac{M^5}{r^5}  - \frac{6957813}{980} \frac{M^6}{r^6} \right. \nonumber \\ 
 && \left.
 + \frac{6488861}{980} \frac{M^7}{r^7} +\frac{667071}{70} \frac{M^8}{r^8} +\frac{15984}{5} \frac{M^9}{r^9} \right) \nonumber \\ 
 &&-\frac{1819}{56448} \left(1+\frac{M}{r}  +\frac{51806}{12733} \frac{M^2}{r^2} +\frac{135383}{63665} \frac{M^3}{r^3}  + \frac{309664}{38199} \frac{M^4}{r^4} - \frac{36264049 }{381990} \frac{M^5}{r^5} - \frac{7873793}{38199} \frac{M^6}{r^6}\right. \nonumber \\ 
&& \left. - \frac{32533551}{63665} \frac{M^7}{r^7} + \frac{73025558 }{63665} \frac{M^8}{r^8} - \frac{8708988}{12733} \frac{M^9}{r^9} - \frac{433800}{1819} \frac{M^{10}}{r^{10}} - \frac{3483648}{1819} \frac{M^{11}}{r^{11}}\right)\left(3 \cos ^2(\theta )-1\right)
 \nonumber \\ 
&& -  \frac{701429}{23708160} \frac{M^2}{r^2} \left( 1 + \frac{1013451}{701429} \frac{M}{r}  + \frac{1154835}{701429} \frac{M^2}{r^2} - \frac{3346744}{701429} \frac{M^3}{r^3} + \frac{3992148}{701429} \frac{M^4}{r^4}   - \frac{9591516}{701429} \frac{M^5}{r^5}\right. \nonumber \\ 
&& \left.+ \frac{94091244}{701429} \frac{M^6}{r^6} - \frac{103967604}{701429} \frac{M^7}{r^7} - \frac{41345640}{701429} \frac{M^8}{r^8} + \frac{109734912}{701429} \frac{M^9}{r^9}\right)  \left( 35 \cos ^4(\theta )-30 \cos ^2(\theta )+3 \right) \Bigg],\\
\delta g_{rr}^\mathrm{CS} =&~~~~ -\zeta \chi^2&\frac{1}{f^2} \frac{M^3}{r^3} \Bigg[ \frac{25}{384} \frac{M}{r} \left( 1 + 3\frac{M}{r} + \frac{322}{5} \frac{M^2}{r^2} \right.  \left. + \frac{198}{5} \frac{M^3}{r^3} + \frac{6276}{175} \frac{M^4}{r^4} - \frac{17496}{25} \frac{M^5}{r^5}   \right) \nonumber \\ 
& & -\frac{201}{1792}  f \left( 1+ \frac{1459}{603} \frac{M}{r} +\frac{20000}{4221} \frac{M^2}{r^2} +\frac{51580}{1407} \frac{M^3}{r^3}  -\frac{7580}{201} \frac{M^4}{r^4} - \frac{22492}{201} \frac{M^5}{r^5}  - \frac{40320}{67} \frac{M^6}{r^6} \right) (3 \cos^2 \theta -1)  \Bigg] \nonumber\\
& +\zeta \chi^4 & \frac{1}{f}\frac{M^3}{r^3}\Bigg[ \frac{5}{384} \frac{M}{r}\frac{1}{f^2}  \left(1+\frac{577 }{175} \frac{M}{r}+\frac{8113}{1200} \frac{M^2}{r^2}-\frac{109309}{11760} \frac{M^3}{r^3}  +\frac{2125311}{3920} \frac{M^4}{r^4} +\frac{267403}{1470 } \frac{M^5}{r^5} + \frac{2001821}{420} \frac{M^6}{r^6} \right.  \nonumber \\ 
& & \left.- \frac{19927289}{980} \frac{M^7}{r^7} +\frac{22161021}{980} \frac{M^8}{r^8} -\frac{12553726}{245} \frac{M^9}{r^9} + \frac{249993}{5} \frac{M^{10}}{r^{10}} + \frac{735264}{25} \frac{M^{11}}{r^{11}}  \right) \nonumber \\ 
& & - \frac{1819}{56448} \frac{1}{f}  \left(1+\frac{1084}{1819} \frac{M}{r}+\frac{62621}{12733 } \frac{M^2}{r^2}+\frac{56726}{1819} \frac{M^3}{r^3}  - \frac{2116475}{38199 } \frac{M^4}{r^4} - \frac{112168723}{381990} \frac{M^5}{r^5} - \frac{36858343}{190995} \frac{M^6}{r^6} \right.  \nonumber \\ 
& & \left.+ \frac{3546621}{9095} \frac{M^7}{r^7} - \frac{47131846}{63665} \frac{M^8}{r^8} -\frac{5777844 }{12733 } \frac{M^9}{r^9} - \frac{32693976}{1819 } \frac{M^{10}}{r^{10}} + \frac{80123904}{1819 } \frac{M^{11}}{r^{11}}  \right)  \left(3 \cos ^2(\theta )-1\right)\nonumber \\ 
& & -  \frac{94699}{4741632} \frac{M^2}{r^2} \left( 1 - \frac{916004}{473495 } \frac{M}{r}  + \frac{2411573}{473495} \frac{M^2}{r^2} + \frac{16109646}{473495} \frac{M^3}{r^3} + \frac{2585472}{43045} \frac{M^4}{r^4} - \frac{23898480}{94699} \frac{M^5}{r^5}\right.  \nonumber \\  
& & \left.  - \frac{314374068}{473495} \frac{M^6}{r^6} - \frac{41960268}{43045} \frac{M^7}{r^7} - \frac{1261951488}{473495 } \frac{M^8}{r^8} \right)  \left( 35 \cos ^4(\theta )-30 \cos ^2(\theta )+3 \right)  \Bigg], \\
\delta g_{\theta\theta}^\mathrm{CS} =& \zeta \chi^2& \frac{201}{1792}M^2 \frac{M}{r}  \left( 1 + \frac{1420}{603} \frac{M}{r} + \frac{18908}{4221} \frac{M^2}{r^2} + \frac{1480}{603} \frac{M^3}{r^3} + \frac{22460}{1407} \frac{M^4}{r^4} + \frac{3848}{201} \frac{M^5}{r^5} + \frac{5376}{67} \frac{M^6}{r^6} \right) (3 \cos^2 \theta -1)\nonumber\\
 +&\zeta \chi^4 &M^2\frac{M}{r}  \Bigg[ \frac{67}{2240} \frac{M^2}{r^2}  \left(1+\frac{104533}{19296} \frac{M}{r}+\frac{583357}{45024} \frac{M^2}{r^2}+\frac{311763}{7504} \frac{M^3}{r^3}  +\frac{3112171}{33768} \frac{M^4}{r^4} +\frac{24899}{1608} \frac{M^5}{r^5}\right.  \nonumber \\ 
& & \left. - \frac{2538845}{11256} \frac{M^6}{r^6}- \frac{190101}{268} \frac{M^7}{r^7} -\frac{18648}{67} \frac{M^8}{r^8} \right) -
  \nonumber \\ 
& & \frac{1819}{56448}  \left(1+\frac{17455}{7276}  \frac{M}{r}   +\frac{148755}{25466} \frac{M^2}{r^2} +\frac{52999}{3638 } \frac{M^3}{r^3} + \frac{3438929}{76398} \frac{M^4}{r^4}+ \frac{5163387}{63665} \frac{M^5}{r^5}\right.  \nonumber \\ 
& &  \left.  + \frac{14491811}{190995} \frac{M^6}{r^6} - \frac{5632}{85} \frac{M^7}{r^7} + \frac{6094488}{12733} \frac{M^8}{r^8} + \frac{1232136}{1819} \frac{M^9}{r^9} \frac{3483648}{1819} \frac{M^{10}}{r^{10}} \right)\left(3 \cos ^2(\theta )-1\right) \nonumber \\ 
& & -  \frac{94699}{4741632}  \frac{M^2}{r^2} \left( 1 + \frac{2984191}{1420485 } \frac{M}{r}  + \frac{2339824}{473495} \frac{M^2}{r^2} + \frac{94116}{8609} \frac{M^3}{r^3} + \frac{45539276}{1420485} \frac{M^4}{r^4} + \frac{16610916}{473495} \frac{M^5}{r^5}\right.  \nonumber \\  
& & \left.  + \frac{7853220}{94699} \frac{M^6}{r^6} - \frac{31810968}{473495} \frac{M^7}{r^7} - \frac{109734912}{473495} \frac{M^8}{r^8} \right)  \left( 35 \cos ^4(\theta )-30 \cos ^2(\theta )+3 \right) \Bigg], \\
\delta g_{\phi\phi}^\mathrm{CS}=& \zeta \chi^2&  \frac{201}{1792} M^2\frac{M}{r}  \left( 1 + \frac{1420}{603} \frac{M}{r} + \frac{18908}{4221} \frac{M^2}{r^2} + \frac{1480}{603} \frac{M^3}{r^3} + \frac{22460}{1407} \frac{M^4}{r^4} + \frac{3848}{201} \frac{M^5}{r^5} + \frac{5376}{67} \frac{M^6}{r^6} \right) (3 \cos^2 \theta -1)\sin^2\theta\nonumber\\
+&\zeta\chi^4 & M^2\frac{M}{r}\Bigg[ \frac{1819}{211680}  \left(1+\frac{17455}{7276} \frac{M}{r}+\frac{106545}{25466} \frac{M^2}{r^2}+\frac{571331}{58208} \frac{M^3}{r^3} +\frac{47090579}{1222368} \frac{M^4}{r^4} +\frac{134570577}{1018640} \frac{M^5}{r^5} + \frac{315848443}{1527960} \frac{M^6}{r^6} \right.  \nonumber \\ 
& & \left. - \frac{118918819}{509320} \frac{M^7}{r^7} -\frac{52589025}{101864} \frac{M^8}{r^8} -\frac{11692683}{7276} \frac{M^9}{r^9}  + \frac{2308824}{1819} \frac{M^{10}}{r^{10}} \right)  \nonumber \\ 
& &  - \frac{9095}{592704 } \left(1+\frac{17455}{7276}  \frac{M}{r}  +\frac{740737}{272850} \frac{M^2}{r^2} + \frac{50790941}{13096800} \frac{M^3}{r^3}  + \frac{100159603}{4365600} \frac{M^4}{r^4}+ \frac{47638909}{727600} \frac{M^5}{r^5}\right.  \nonumber \\ 
& &  \left.  + \frac{30022409}{654840} \frac{M^6}{r^6} - \frac{84837063}{363800} \frac{M^7}{r^7} - \frac{5835501}{72760} \frac{M^8}{r^8} - \frac{29787597}{181900} \frac{M^9}{r^9}  + \frac{89318376}{45475} \frac{M^{10}}{r^{10}} \right)\left(3 \cos ^2(\theta )-1\right)  \nonumber \\
& &  + \frac{1819}{658560}  \left( 1 + \frac{17455}{7276} \frac{M}{r} - \frac{198514}{180081} \frac{M^2}{r^2}    - \frac{4120646}{540243} \frac{M^3}{r^3} - \frac{1309801}{360162} \frac{M^4}{r^4} + \frac{10009}{2805} \frac{M^5}{r^5}  - \frac{294356903}{2701215} \frac{M^6}{r^6} \right. \nonumber \\  
& & \left. - \frac{109195222}{300135} \frac{M^7}{r^7} - \frac{4348890}{20009} \frac{M^8}{r^8} +\frac{7512372}{20009} \frac{M^9}{r^9} +\frac{54867456}{20009} \frac{M^{10}}{r^{10}}  \right) \left( 35 \cos ^4(\theta )-30 \cos ^2(\theta )+3 \right)  \nonumber \\  
& &  + \frac{701429}{156473856} \frac{M^2}{r^2} \left( 1 + \frac{5962075}{2104287} \frac{M}{r}   + \frac{4434376}{701429} \frac{M^2}{r^2}   + \frac{7777884}{701429} \frac{M^3}{r^3} + \frac{59811476}{2104287 } \frac{M^4}{r^4} + \frac{28251588}{701429} \frac{M^5}{r^5}\right.  \nonumber \\  
& & \left. + \frac{66282516}{701429} \frac{M^6}{r^6} + \frac{4767336}{701429} \frac{M^7}{r^7} - \frac{109734912}{701429} \frac{M^8}{r^8} \right)  \left(231 \cos ^6(\theta )-315 \cos ^4(\theta )+105 \cos ^2(\theta )-5 \right) \Bigg] ,\\
g_{t\phi}^\mathrm{CS} =& 
~~~~\zeta \chi& \frac{5}{8}M\frac{M^4}{r^4} \left( 1+\frac{12}{7}\frac{M}{r} + \frac{27}{10} \frac{M^2}{r^2} \right) \sin^2 (\theta)\nonumber\\
&+\zeta \chi^3& M\frac{M^3}{r^3}  \Bigg[ -\frac{8819}{141120} \left(1+\frac{60155}{35276} \frac{M}{r}+\frac{8545}{8819} \frac{M^2}{r^2}-\frac{19828}{26457} \frac{M^3}{r^3} +\frac{563669}{26457} \frac{M^4}{r^4} +\frac{549630}{8819} \frac{M^5}{r^5} + \frac{873180}{8819} \frac{M^6}{r^6} \right)  \nonumber \\ 
& &  - \frac{8819}{56448}  \left(1+\frac{24875}{35276}  \frac{M}{r}  +\frac{95}{17638} \frac{M^2}{r^2}  + \frac{90188}{26457} \frac{M^3}{r^3}  + \frac{684818}{26457} \frac{M^4}{r^4}+ \frac{385542}{8819} \frac{M^5}{r^5} \right.  \nonumber \\ 
& &  \left.+ \frac{418572}{8819} \frac{M^6}{r^6}  - \frac{508032}{8819} \frac{M^7}{r^7} \right)\left(3 \cos ^2(\theta )-1\right)\Bigg] \sin ^2(\theta) \nonumber \\ 
&  + \zeta \chi^5&  M\frac{M^3}{r^3} \Bigg[ \frac{3840911}{142248960}  \left(1+\frac{3368875}{7681822} \frac{M}{r}+\frac{539981961}{211250105} \frac{M^2}{r^2} \right.  \nonumber \\ 
& & \left. +\frac{63963088}{211250105} \frac{M^3}{r^3}  - \frac{28203665}{84500042 } \frac{M^4}{r^4} -\frac{218979789}{84500042} \frac{M^5}{r^5}  + \frac{6554146711}{42250021} \frac{M^6}{r^6}  + \frac{1870270010}{3840911} \frac{M^7}{r^7} + \frac{3798260802}{3840911} \frac{M^8}{r^8} \right. \nonumber \\ 
& &  \left.  -\frac{1514962386}{3840911} \frac{M^9}{r^9}  - \frac{2505947220}{3840911} \frac{M^{10}}{r^{10}}  - \frac{1184222592}{3840911} \frac{M^{11}}{r^{11}} \right)  \nonumber \\ 
& &  + \frac{3840911}{56899584}  \left(1+\frac{10036795}{7681822}  \frac{M}{r}  +\frac{949643961}{211250105} \frac{M^2}{r^2}  + \frac{1196741284}{211250105} \frac{M^3}{r^3}  + \frac{304195064}{42250021} \frac{M^4}{r^4} \right. \nonumber \\ 
& & \left.+ \frac{646950168}{211250105} \frac{M^5}{r^5}+ \frac{19300145456}{211250105} \frac{M^6}{r^6} + \frac{1091987984}{3840911} \frac{M^7}{r^7} + \frac{13626382752}{19204555} \frac{M^8}{r^8} \right.  \nonumber \\ 
& &  \left. - \frac{914366016}{3840911} \frac{M^9}{r^9}  -  \frac{290957184}{3840911} \frac{M^{10}}{r^{10}}  -  \frac{3511517184}{3840911} \frac{M^{11}}{r^{11}}  \right) \left(3 \cos ^2(\theta )-1\right) \nonumber \\ 
& & + \frac{65029949}{1738598400} \frac{M^2}{r^2}  \left( 1 + \frac{247489546}{195089847} \frac{M}{r}   - \frac{192857740}{585269541} \frac{M^2}{r^2}  + \frac{201416960}{195089847} \frac{M^3}{r^3} + \frac{6952033840}{195089847} \frac{M^4}{r^4} \right.  \nonumber \\  
& & \left. + \frac{49673623120}{585269541} \frac{M^5}{r^5} + \frac{8477276720}{65029949} \frac{M^6}{r^6}  - \frac{7984872720}{65029949} \frac{M^7}{r^7}  \right.  \nonumber \\ 
& & \left. - \frac{5714422560}{65029949 } \frac{M^8}{r^8} +\frac{8047226880}{65029949} \frac{M^9}{r^9}   \right)\left( 35 \cos ^4(\theta )-30 \cos ^2(\theta )+3 \right) \Bigg]\sin ^2(\theta ) \,.
\end{eqnarray}
\end{widetext}
\subsection{Scalar Gauss-Bonnet Gravity}
Presented below is the slow-rotation, small-coupling BH solution of sGB gravity in Boyer-Lindquist coordinates
\begin{align}
    g_{\alpha\beta}^\mathrm{GB} = g_{\alpha\beta}^{\mathrm{GR}} + \delta g_{\alpha\beta}^\mathrm{GB}. 
\end{align}
The spherically symmetric term of the sGB correction $\delta g_{\alpha\beta}^\mathrm{GB}$ given below was computed by \cite{Yunes_2011} and is valid for any quadratic gravity theory with a scalar coupled to the Kretschmann scalar $R_{\alpha\beta\gamma\delta}R^{\alpha\beta\gamma\delta}$ through a linear coupling function, as is the linear-in-spin term derived by \cite{Pani_2011}. The quadratic-in-spin~\cite{Ayzenberg_2014} and higher order terms~\cite{Maselli:2015tta} were derived in the context of Einstein-dilaton-Gauss-Bonnet gravity under the approximation that only the linear term in a small-$\vartheta$ expansion of the dilatonic coupling function need be considered. 
\begin{widetext}
\begin{eqnarray}
\delta g_{tt}^\mathrm{GB}=&~-\zeta&\frac{1}{3} \frac{M^3}{r^3}\left(1
+26\frac{M}{r}
+\frac{66}{5}\frac{M^2}{r^2}
+\frac{96}{5}\frac{M^3}{r^3}
-80\frac{M^2}{r^2}\right)\nonumber\\
&+\zeta\chi^2&\frac{175}{1046}\frac{M^3}{r^3}\Bigg[\left(
1
+14\frac{M}{r}
+\frac{52}{5}\frac{M^2}{r^2}
+\frac{1214}{15}\frac{M^3}{r^3}
+68\frac{M^4}{r^4}
+\frac{724}{5}\frac{M^5}{r^5}
-\frac{11264}{15}\frac{M^6}{r^6}
+\frac{160}{3}\frac{M^7}{r^7}\right)\nonumber\\
&&-\frac{8926}{875}\left(
1
+\frac{M}{r}
+\frac{27479}{31241}\frac{M^2}{r^2}
-\frac{2275145}{187446}\frac{M^3}{r^3}
-\frac{2030855}{93723}\frac{M^4}{r^4}
-\frac{99975}{4463}\frac{M^5}{r^5}
+\frac{1128850}{13389}\frac{M^6}{r^6}\right.\nonumber\\&&\left.
+\frac{194600}{4463}\frac{M^7}{r^7}
-\frac{210000}{4463}\frac{M^8}{r^8}\right)\left(3\cos^2(\theta)-1\right)\Bigg]\nonumber\\
~&~+ \zeta\chi^4&\frac{M^{3}}{r^{3}}\Bigg[\frac{1}{16} \left(1
+18\frac{M}{r}
+{\frac{26}{3}}\frac{M^{2}}{r^{2}}
-{\frac{1670264}{39375}}\frac{M^{3}}{r^{3}}
-{\frac{3770006}{275625}}\frac{M^{4}}{r^{4}}
-{\frac{2998312}{5625}}\frac{M^{5}}{r^{5}}
-{\frac{149817242}{118125}}\frac{M^{6}}{r^{6}}\right.\nonumber\\&&\left.
-{\frac{2140167016}{826875}}\frac{M^{7}}{r^{7}}
+{\frac{411692276}{55125}}\frac{M^{8}}{r^{8}}
+{\frac{23175976}{33075}}\frac{M^{9}}{r^{9}}
-{\frac{417248}{189}}\frac{M^{10}}{r^{10}}
-{\frac{528832}{75}}\frac{M^{11}}{r^{11}}
-{\frac{199936}{15}}\frac{M^{12}}{r^{12}}\right)\nonumber\\
&&+\frac{33863}{68600} \left( 1
+\frac{M}{r}
+{\frac{12566236}{3555615}}\frac{M^{2}}{r^{2}}
-{\frac{41599844}{10666845}}\frac{M^{3}}{r^{3}}
-{\frac{48084604}{10666845}}\frac{M^{4}}{r^{4}}
-{\frac{161984624}{1523835}}\frac{M^{5}}{r^{5}}\right.\nonumber\\&&\left.
-{\frac{932560088}{4571505}}\frac{M^{6}}{r^{6}}
-{\frac{1863420316}{4571505}}\frac{M^{7}}{r^{7}}
+{\frac{444019304}{304767}}\frac{M^{8}}{r^{8}}
+{\frac{214419560}{914301}}\frac{M^{9}}{r^{9}}
-{\frac{6551440}{914301}}\frac{M^{10}}{r^{10}}\right.\nonumber\\&&\left.
+{\frac{93343040}{101589}}\frac{M^{11}}{r^{11}}
-{\frac{87651200}{33863}}\frac{M^{12}}{r^{12}}\right) \left( 3 \cos^{2}(\theta) -1 \right)\nonumber\\
&&+\frac{1688837
}{3601500} \frac{M^{2}}{r^{2}}\left( 
1
+{\frac{12733507}{8444185}}\frac{M}{r}
+{\frac{3018832}{1688837}}\frac{M^{2}}{r^{2}}
-{\frac{503900579}{50665110}}\frac{M^{3}}{r^{3}}
-{\frac{728918617}{25332555}}\frac{M^{4}}{r^{4}}\right.\nonumber\\&&\left.
-{\frac{355747119}{8444185}}\frac{M^{5}}{r^{5}}
+{\frac{312047848}{1688837}}\frac{M^{6}}{r^{6}}
+{\frac{121681000}{1688837}}\frac{M^{7}}{r^{7}}
-{\frac{127013880}{1688837}}\frac{M^{8}}{r^{8}}\right.\nonumber\\&&\left.
-{\frac{190653120}{1688837}}\frac{M^{9}}{r^{9}}
-{\frac{85612800}{1688837}}\frac{M^{10}}{r^{10}} \right)\left( 35 \cos(\theta) ^{4}-30 \cos^{2}(\theta) + 3\right),\\
\delta g_{rr}^\mathrm{GB}=&~ -\zeta& \frac{1}{f^2}\frac{M^2}{r^2}\left(
1
+\frac{M}{r}
+\frac{52}{3}\frac{M^2}{r^2}
+2\frac{M^3}{r^3}
+\frac{16}{5}\frac{M^2}{r^2}
-\frac{368}{3}\frac{M^2}{r^2}\right)\nonumber\\
&+\zeta \chi^2 & \frac{1}{2}\frac{1}{f^3}\frac{M^2}{r^2}\Bigg[\left(1
-\frac{M}{r}
+\frac{10M^2}{r^2}
-\frac{12M^3}{r^3}
+\frac{218}{3}\frac{M^4}{r^4}
+\frac{128}{3}\frac{M^5}{r^5}
-\frac{724}{15}\frac{M^6}{r^6}
-\frac{22664}{15}\frac{M^7}{r^7}
+\frac{25312}{15}\frac{M^8}{r^8}
+\frac{1600}{3}\frac{M^9}{r^9}\right)\nonumber\\&&
-\frac{8932}{2625}\frac{M}{r}\left(
1
-\frac{5338}{4463}\frac{M}{r}
-\frac{59503}{31241}\frac{M^2}{r^2}
-\frac{7433843}{187446}\frac{M^3}{r^3}
+\frac{13462040}{93723}\frac{M^4}{r^4}
-\frac{7072405}{31241}\frac{M^5}{r^5}\right.\nonumber \\&&
\left.+\frac{9896300}{13389}\frac{M^6}{r^6}
-\frac{28857700}{13389}\frac{M^7}{r^7}
+\frac{13188000}{4463}\frac{M^8}{r^8}
-\frac{7140000}{4463}\frac{M^9}{r^9}\right)\left(3\cos^2(\theta)-1\right)\Bigg]\nonumber\\
~&~+ \zeta\chi^4&\frac{1}{{f}}\frac{M^{2}}{r^{2}}\Bigg[
\frac{3}{16}\frac{1}{{f}^{3}} \left( 
1
-3\frac{M}{r}
+{\frac{88}{9}}\frac{M^{2}}{r^{2}}
-{\frac{5194382}{118125}}\frac{M^{3}}{r^{3}}
+{\frac{3312142}{118125}}\frac{M^{4}}{r^{4}}
-{\frac{20190122}{826875}}\frac{M^{5}}{r^{5}}
-{\frac{2284837154}{2480625}}\frac{M^{6}}{r^{6}}\right.\nonumber \\&&
\left.
+{\frac{3309895118}{2480625}}\frac{M^{7}}{r^{7}}
+{\frac{7046059012}{2480625}}\frac{M^{8}}{r^{8}}
+{\frac{26212989676}{2480625}}\frac{M^{9}}{r^{9}}
-{\frac{24838664888}{496125}}\frac{M^{10}}{r^{10}}
+{\frac{1768104368}{70875}}\frac{M^{11}}{r^{11}}\right.\nonumber \\&&
\left.
+{\frac{285023776}{3675}}\frac{M^{12}}{r^{12}}
+{\frac{5070208}{135}}\frac{M^{13}}{r^{13}}
-{\frac{103298816}{225}}\frac{M^{14}}{r^{14}}
+{\frac{3998720}{9}}\frac{M^{15}}{r^{15}} \right)\nonumber\\
&&+\frac{33863 }{68600}\frac{1}{{f}^{2}} \frac{M}{r}\left( 
1
-{\frac{135889}{101589}}\frac{M}{r}
+{\frac{19453978}{10666845}}\frac{M^{2}}{r^{2}}
-{\frac{29250392}{3555615}}\frac{M^{3}}{r^{3}}
+{\frac{567359132}{10666845}}\frac{M^{4}}{r^{4}}
-{\frac{4647986792}{32000535}}\frac{M^{5}}{r^{5}}\right.\nonumber \\&&
\left.
-{\frac{3384415304}{4571505}}\frac{M^{6}}{r^{6}}
+{\frac{82733044}{169315}}\frac{M^{7}}{r^{7}}
+{\frac{7096592768}{914301}}\frac{M^{8}}{r^{8}}
-{\frac{7009960792}{914301}}\frac{M^{9}}{r^{9}}
-{\frac{55778080}{33863}}\frac{M^{10}}{r^{10}}\right.\nonumber \\&&
\left.
-{\frac{19265271200}{304767}}\frac{M^{11}}{r^{11}}
+{\frac{17704413440}{101589}}\frac{M^{12}}{r^{12}}
-{\frac{4382560000}{33863}}\frac{M^{13}}{r^{13}} \right)
\left( 3 \cos^{2}(\theta) -1 \right)\nonumber\\
&&+\frac{5819941}{18007500}\frac{M^{3}}{r^{3}}\left( 1
-{\frac{10138655}{5819941}}\frac{M}{r}
+{\frac{24493721}{5819941}}\frac{M^{2}}{r^{2}}
+{\frac{70965055}{34919646}}\frac{M^{3}}{r^{3}}
-{\frac{69839000}{5819941}}\frac{M^{4}}{r^{4}}\right.\nonumber\\&&
\left.
+{\frac{113406265}{5819941}}\frac{M^{5}}{r^{5}}
+{\frac{1452929450}{5819941}}\frac{M^{6}}{r^{6}}
-{\frac{1313724300}{5819941}}\frac{M^{7}}{r^{7}}
-{\frac{10855538400}{5819941}}\frac{M^{8}}{r^{8}}\right.\nonumber\\&&
\left.
-{\frac{5350800000}{5819941}}\frac{M^{9}}{r^{9}}\right)\left(35 \cos(\theta) ^{4}-30 \cos^{2}(\theta)  +3\right)\Bigg],\\
\delta g_{\theta\theta}^\mathrm{GB}=&-\zeta\chi^2&\frac{4463}{2625}M^2 \frac{M}{r}\left(
1
+\frac{10370}{4463}\frac{M}{r}
+\frac{266911}{62482}\frac{M^2}{r^2}
+\frac{63365}{13389}\frac{M^3}{r^3}
-\frac{309275}{31241}\frac{M^4}{r^4}
-\frac{81350}{4463}\frac{M^5}{r^5}\right.\nonumber \\&&\left.
-\frac{443800}{13389}\frac{M^6}{r^6}
+\frac{210000}{4463}\frac{M^7}{r^7}\right) \left(3\cos^2\theta-1\right)\nonumber\\
~&~- \zeta\chi^4&{M}^{2}\frac{M}{r}\Bigg[\frac{1}{12}\frac{M}{r} \left( 
1
+{\frac{123908}{13125}}\frac{M}{r}
+{\frac{1079443}{26250}}\frac{M^{2}}{r^{2}}
+{\frac{2876786}{30625}}\frac{M^{3}}{r^{3}}
+{\frac{21064837}{551250}}\frac{M^{4}}{r^{4}}
-{\frac{28148774}{275625}}\frac{M^{5}}{r^{5}}\right.\nonumber \\&&
\left.
+{\frac{63849}{875}}\frac{M^{6}}{r^{6}}
+{\frac{1740434}{11025}}\frac{M^{7}}{r^{7}}
-{\frac{697976}{1575}}\frac{M^{8}}{r^{8}}
-{\frac{100528}{25}}\frac{M^{9}}{r^{9}}
-{\frac{49984}{5}}\frac{M^{10}}{r^{10}} \right)\nonumber\\
&&-\frac{33863}{68600}\left( 
1
+{\frac{161475}{67726}}\frac{M}{r}
+{\frac{57983182}{10666845}}\frac{M^{2}}{r^{2}}
+{\frac{4243970}{304767}}\frac{M^{3}}{r^{3}}
+{\frac{296394592}{10666845}}\frac{M^{4}}{r^{4}}
+{\frac{100526248}{4571505}}\frac{M^{5}}{r^{5}}\right.\nonumber \\&&
\left.
+{\frac{55402796}{4571505}}\frac{M^{6}}{r^{6}}
-{\frac{16491328}{101589}}\frac{M^{7}}{r^{7}}
-{\frac{118487080}{914301}}\frac{M^{8}}{r^{8}}
-{\frac{269150000}{914301}}\frac{M^{9}}{r^{9}}\right.\nonumber \\&&
\left.
-{\frac{56651840}{101589}}\frac{M^{10}}{r^{10}}
+{\frac{87651200}{33863}}\frac{M^{11}}{r^{11}} \right)
\left( 3 \cos^{2}(\theta) -1 \right)\nonumber\\
&&-\frac{5819941}{18007500} \frac{M^{2}}{r^{2}}  \left( 
1
+{\frac{37902886}{17459823}}\frac{M}{r}
+{\frac{57133373}{11639882}}\frac{M^{2}}{r^{2}}
+{\frac{96691049}{17459823}}\frac{M^{3}}{r^{3}}
-{\frac{329394989}{52379469}}\frac{M^{4}}{r^{4}}\right.\nonumber \\&&
\left.
-{\frac{185345440}{5819941}}\frac{M^{5}}{r^{5}}
+{\frac{338079000}{5819941}}\frac{M^{6}}{r^{6}}
+{\frac{1042827800}{5819941}}\frac{M^{7}}{r^{7}}
+{\frac{1216689600}{5819941}}\frac{M^{8}}{r^{8}}\right.\nonumber \\&&
\left.
+{\frac{428064000}{5819941}}\frac{M^{9}}{r^{9}}\right)\left(35 \cos(\theta) ^{4}-30 \cos^{2}(\theta) +3 \right)\Bigg],\\
\delta g_{\phi\phi}^\mathrm{GB}=&-\zeta\chi^2&\frac{4463}{2625}M^2 \frac{M}{r}\left(
1
+\frac{10370}{4463}\frac{M}{r}
+\frac{266911}{62482}\frac{M^2}{r^2}
+\frac{63365}{13389}\frac{M^3}{r^3}
-\frac{309275}{31241}\frac{M^4}{r^4}
-\frac{81350}{4463}\frac{M^5}{r^5}\right.\nonumber \\&&\left.
-\frac{443800}{13389}\frac{M^6}{r^6}
+\frac{210000}{4463}\frac{M^7}{r^7}\right) \left(3\cos^2(\theta)-1\right)\sin^2(\theta)\nonumber\\
~&~- \zeta\chi^4&{M}^{2} \frac{M}{r}\Bigg[-\frac{1}{12}{M}^{2} \frac{M^{2}}{r^{2}} \left( 
1
+4\frac{M}{r}
+{\frac{503971}{26250}}\frac{M^{2}}{r^{2}}
+{\frac{346746}{6125}}\frac{M^{3}}{r^{3}}
+{\frac{1085069}{61250}}\frac{M^{4}}{r^{4}}
+{\frac{7739222}{91875}}\frac{M^{5}}{r^{5}}
+{\frac{3391903}{6125}}\frac{M^{6}}{r^{6}}\right.\nonumber \\&&
\left.
+{\frac{489746}{1225}}\frac{M^{7}}{r^{7}}
-{\frac{771752}{525}}\frac{M^{8}}{r^{8}}
-{\frac{157488}{25}}\frac{M^{9}}{r^{9}}
-{\frac{49984}{5}}\frac{M^{10}}{r^{10}} \right)\nonumber\\
&&-\frac{33863}{68600}\left(1
+{\frac{161475}{67726}}\frac{M}{r}
+{\frac{13496074}{3555615}}\frac{M^{2}}{r^{2}}
+{\frac{2053418}{304767}}\frac{M^{3}}{r^{3}}
+{\frac{133155712}{10666845}}\frac{M^{4}}{r^{4}}
-{\frac{491696}{169315}}\frac{M^{5}}{r^{5}}
+{\frac{2648332}{1523835}}\frac{M^{6}}{r^{6}}\right.\nonumber \\&&
\left.
-{\frac{11833088}{101589}}\frac{M^{7}}{r^{7}}
-{\frac{243560}{33863}}\frac{M^{8}}{r^{8}}
-{\frac{59307920}{304767}}\frac{M^{9}}{r^{9}}
-{\frac{61512640}{101589}}\frac{M^{10}}{r^{10}}
+{\frac{87651200}{33863}}\frac{M^{11}}{r^{11}} \right) \left( 3 \cos^{2}(\theta) -1 \right)\nonumber\\
&&-\frac{1688837}{3601500} \frac{M^{2}}{r^{2}}\left( 1
+{\frac{14696906}{5066511}}\frac{M}{r}
+{\frac{111147137}{16888370}}\frac{M^{2}}{r^{2}}
+{\frac{251117741}{25332555}}\frac{M^{3}}{r^{3}}
+{\frac{192852331}{75997665}}\frac{M^{4}}{r^{4}}\right.\nonumber \\&&
\left.
-{\frac{39411008}{1688837}}\frac{M^{5}}{r^{5}}
+{\frac{22290800}{1688837}}\frac{M^{6}}{r^{6}}
+{\frac{125696760}{1688837}}\frac{M^{7}}{r^{7}}
+{\frac{179128320}{1688837}}\frac{M^{8}}{r^{8}}\right.\nonumber \\&&
\left.
+{\frac{85612800}{1688837}}\frac{M^{9}}{r^{9}}\right)\left( 35 \cos(\theta) ^{4}-30 \cos^{2}(\theta) +3 \right)\Bigg]\sin^{2}(\theta),\\
\delta g_{t\phi}^\mathrm{GB}=&~ \zeta\chi&\frac{3}{5}M \frac{M^3}{r^3}\left(1+\frac{140}{9}\frac{M}{r}+10\frac{M^2}{r^2}+16\frac{M^3}{r^3}+\frac{400}{9}\frac{M^2}{r^2}-\frac{368}{3}\frac{M^2}{r^2}\right)\sin^2\theta\nonumber\\
~&~ +\zeta\chi^3&M\frac{M^{3}}{r^{3}}\Bigg[\frac{71933}{110250} \left( 
1
-{\frac{218926}{71933}}\frac{M}{r}
-{\frac{343372}{71933}}\frac{M^{2}}{r^{2}}
-{\frac{1322138}{215799}}\frac{M^{3}}{r^{3}}
+{\frac{3798100}{215799}}\frac{M^{4}}{r^{4}}
+{\frac{3721500}{71933}}\frac{M^{5}}{r^{5}}\right.\nonumber \\&&
\left.
+{\frac{12857600}{71933}}\frac{M^{6}}{r^{6}}
+{\frac{980000}{71933}}\frac{M^{7}}{r^{7}} \right)\nonumber\\
&&+\frac{26252}{11025}\left( 
1
+{\frac{37537}{52504}}\frac{M}{r}
-{\frac{631}{6563}}\frac{M^{2}}{r^{2}}
-{\frac{3620183}{787560}}\frac{M^{3}}{r^{3}}
-{\frac{441053}{78756}}\frac{M^{4}}{r^{4}}
+{\frac{218325}{26252}}\frac{M^{5}}{r^{5}}
+{\frac{743155}{13126}}\frac{M^{6}}{r^{6}}\right.\nonumber \\&&
\left.
+{\frac{253330}{6563}}\frac{M^{7}}{r^{7}}
-{\frac{220500}{6563}}\frac{M^{8}}{r^{8}}\right)\left( 3 \cos^{2}(\theta) -1 \right)\Bigg]\sin^{2}(\theta)\nonumber\\
~&~ -\zeta\chi^5&M\frac{M^{3}}{r^{3}}\Bigg[\frac{5214311}{11113200} \left( 1
+{\frac{18315884}{5214311}}\frac{M}{r}
+{\frac{1285910646}{286787105}}\frac{M^{2}}{r^{2}}
-{\frac{83542296}{286787105}}\frac{M^{3}}{r^{3}}
-{\frac{2390176062}{1433935525}}\frac{M^{4}}{r^{4}}\right.\nonumber \\&&
\left.
-{\frac{3315551688}{286787105}}\frac{M^{5}}{r^{5}}
+{\frac{60975599622}{1433935525}}\frac{M^{6}}{r^{6}}
+{\frac{31793661144}{130357775}}\frac{M^{7}}{r^{7}}
+{\frac{25384874508}{26071555}}\frac{M^{8}}{r^{8}}
+{\frac{926303112}{5214311}}\frac{M^{9}}{r^{9}}\right.\nonumber \\&&
\left.
-{\frac{709922976}{5214311}}\frac{M^{10}}{r^{10}}
-{\frac{2550405312}{5214311}}\frac{M^{11}}{r^{11}}
-{\frac{130394880}{73441}}\frac{M^{12}}{r^{12}}\right)\nonumber\\&&
+\frac{991019}{1111320}\left( 1
+{\frac{1106096}{991019}}\frac{M}{r}
+{\frac{1560149043}{381542315}}\frac{M^{2}}{r^{2}}
+{\frac{1787352258}{1907711575}}\frac{M^{3}}{r^{3}}
-{\frac{1452309372}{272530225}}\frac{M^{4}}{r^{4}}\right.\nonumber \\&&
\left.
-{\frac{1511044740}{76308463}}\frac{M^{5}}{r^{5}}
+{\frac{9922741872}{272530225}}\frac{M^{6}}{r^{6}}
+{\frac{5186519532}{24775475}}\frac{M^{7}}{r^{7}}
+{\frac{3997846296}{4955095}}\frac{M^{8}}{r^{8}}
+{\frac{234802872}{991019}}\frac{M^{9}}{r^{9}}\right.\nonumber \\&&
\left.
+{\frac{150454416}{991019}}\frac{M^{10}}{r^{10}}
+{\frac{763148736}{991019}}\frac{M^{11}}{r^{11}}
-{\frac{1419949440}{991019}}\frac{M^{12}}{r^{12}}
\right)\left( 3 \cos^{2}(\theta) -1 \right) \nonumber\\&&
+\frac{144724751}{237699000}\frac{M^{2}}{r^{2}} \left( 1
+{\frac{607208752}{434174253}}\frac{M}{r}
+{\frac{293625416}{2170871265}}\frac{M^{2}}{r^{2}}
-{\frac{2682551041}{723623755}}\frac{M^{3}}{r^{3}}
-{\frac{2588150782}{723623755}}\frac{M^{4}}{r^{4}}\right.\nonumber \\&&
\left.
+{\frac{62304236918}{2170871265}}\frac{M^{5}}{r^{5}}
+{\frac{21405746208}{144724751}}\frac{M^{6}}{r^{6}}
+{\frac{12464889360}{144724751}}\frac{M^{7}}{r^{7}}
-{\frac{3246806640}{144724751}}\frac{M^{8}}{r^{8}}\right.\nonumber \\&&
\left.
-{\frac{10844507520}{144724751}}\frac{M^{9}}{r^{9}}
-{\frac{5650444800}{144724751}}\frac{M^{10}}{r^{10}}\right)\left( 35 \cos(\theta) ^{4}-30 \cos^{2}(\theta)  +3\right)\Bigg] \sin^{2}(\theta).
\end{eqnarray}
\end{widetext}
\section{Lorentz Transformations in the Newman-Penrose Formalism}\label{SEC:LorentzTransformations}
Included for completeness is an overview of Lorentz transformations in Newman-Penrose formalism following the presentation in \cite{Chandrasekhar_1992}. The three classes of transformations that can be performed on the complex null tetrad are detailed below. The transformation parameters $A$ and $B$ are complex and $X$ and $Y$ are real. The 6 real degrees of freedom correspond to the 6 degrees of freedom of the Lorentz group. Each of the transformations preserves the normalization and orthogonality requirements of the tetrad. 

\subsection{Class I}
A transformation of class I leaves  $l^{\alpha}$ and $\Psi_0$ unchanged. Under this class of transformation, the tetrad becomes
\begin{align}
     l \rightarrow&~ l,\nonumber\\
    n \rightarrow&~ n + \bar{A}m +A\bar{m} + A\bar{A}l,\nonumber\\ 
     m \rightarrow&~ m + Al,  \nonumber\\
     \bar{m} \rightarrow&~ \bar{m} + \bar{A}l.
\end{align}
 The corresponding transformations of the Weyl scalars are 
\begin{align}
     \Psi_0 \rightarrow &~\Psi_0,\nonumber\\ 
    \Psi_1 \rightarrow &~\Psi_1 + \bar{A}\Psi_0,\nonumber\\ 
    \Psi_2 \rightarrow &~\Psi_2+ 2\bar{A}\Psi_1 + \bar{A}^2\Psi_0,\nonumber\\ 
    \Psi_3 \rightarrow &~\Psi_3 + 3 \bar{A}\Psi_2+ 3\bar{A}^2\Psi_1 + \bar{A}^3\Psi_0,\nonumber\\ 
    \Psi_4 \rightarrow &~\Psi_4 + 4 \bar{A} \Psi_3 + 6 \bar{A}^2 \Psi_2,+ 4\bar{A}^3\Psi_1 + \bar{A}^4\Psi_0.
\end{align}
\subsection{Class II}
A class II transformation leaves $n^{\alpha}$ and $\Psi_4$ unchanged. Under this class of transformation, the tetrad becomes
\begin{align}
     l \rightarrow &~l + \bar{B}m + B\bar{m} + B\bar{B}n,\nonumber\\
    n \rightarrow&~ n,\nonumber\\
    m \rightarrow&~ m + Bl, \nonumber\\ 
    \bar{m} \rightarrow&~ \bar{m} + \bar{B}l, 
\end{align}
and the Weyl scalars become 
\begin{align}
    \Psi_0 \rightarrow &~\Psi_0 + 4 B \Psi_1 + 6 B ^2 \Psi_2+ 4B^3\Psi_3 + B^4\Psi_4,\nonumber\\ 
    \Psi_1 \rightarrow &~\Psi_1 + 3 B\Psi_2+ 3B^2\Psi_3 + B^3\Psi_4,\nonumber\\ 
    \Psi_2 \rightarrow &~\Psi_2 + 2B\Psi_3 + B^2\Psi_4,\nonumber\\ 
    \Psi_3 \rightarrow &~\Psi_3 + B\Psi_4,\nonumber\\ 
    \Psi_4 \rightarrow &~\Psi_4.
\end{align}
The Petrov type of a spacetime is determined by counting the number of ways $\Psi_0$ can be made to vanish under a class II transformation. Principal null directions of the spacetime are the vectors $l^\alpha$ in the frames where $\Psi_0 =0.$

\subsection{Class III}
A class III transformation preserves the directions of $l^\alpha$ and $n^\alpha$ and rotates $m^\alpha$ and $\bar{m}^\alpha$ in their plane. The transformed tetrad is
\begin{align}
     l \rightarrow&~ l/Y,\nonumber\\
     n \rightarrow&~ Yn,\nonumber\\
     m \rightarrow&~ e^{iX} m,\nonumber\\
     \bar{m} \rightarrow~&~ e^{-iX}\bar{m},
\end{align}
and the transformed Weyl scalars are
\begin{align}
     \Psi_0 \rightarrow&~  Y^{-2}e^{2iX}\Psi_0,\nonumber\\
     \Psi_1  \rightarrow&~ Y^{-1}e^{iX}\Psi_1,\nonumber\\
     \Psi_2 \rightarrow&~  \Psi_2,\nonumber\\
     \Psi_3 \rightarrow&~  Ye^{-iX}\Psi_3,\nonumber\\
     \Psi_4 \rightarrow&~ Y^{2}e^{-2iX}\Psi_4.
\end{align}
\\
\section{Results}\label{SEC:Results}
\subsection{Dynamical Chern-Simons Gravity}
In this appendix we extend the results presented in Sec. \ref{SEC:dCSPNDs}, including dCS corrections through $\mathcal{O}(\chi^2\zeta)$ to the PND, tetrad and Weyl scalars. The full results, which include corrections to $\mathcal{O}(\chi^4\zeta),$ are collected in a \textit{Mathematica} notebook that is provided in the Supplemental Material.

The PNDs of dCS are given by 
\begin{align}
k^\alpha_{1,\mathrm{CS}}=&~k^\alpha_{1,\mathrm{GR}} + \delta k^\alpha_{1,\mathrm{CS}} +\mathcal{O}(\chi^3\sqrt{\zeta}), \nonumber\\
k^\alpha_{2,\mathrm{CS}}=&~k^\alpha_{1,\mathrm{GR}} + \delta k^\alpha_{2,\mathrm{CS}}+\mathcal{O}(\chi^3\sqrt{\zeta}), \nonumber\\
k^\alpha_{3,\mathrm{CS}}=&~k^\alpha_{2,\mathrm{GR}} + \delta k^\alpha_{3,\mathrm{CS}}+\mathcal{O}(\chi^3\sqrt{\zeta}), \nonumber\\
k^\alpha_{4,\mathrm{CS}}=&~k^\alpha_{2,\mathrm{GR}} + \delta k^\alpha_{4,\mathrm{CS}}+\mathcal{O}(\chi^3\sqrt{\zeta}),\nonumber\\
\end{align}
where the $k^\alpha_{1,2,\mathrm{GR}}$ are given in Sec. \ref{SEC:Basics in GR} and 
\begin{align}\delta k^\alpha_{i,\mathrm{CS}} =
k^{\alpha(1,\frac{1}{2})}_{i,\mathrm{CS}} 
+  k^{\alpha(2,\frac{1}{2})}_{i,\mathrm{CS}} 
+ k^{\alpha(1,1)}_{i,\mathrm{CS}} 
+ k^{\alpha(2,1)}_{i,\mathrm{CS}},
\end{align}
with  $ k^{\alpha(m,n)}_{i,\mathrm{CS}}\propto \chi^m\zeta^n$. The terms that make up the nonvanishing components of $\delta k^\alpha_{1\mathrm{CS}}$ are
\begin{widetext}
\begin{eqnarray}
k^{t(2,\frac{1}{2})}_{1,\mathrm{CS}}=
 &~\sqrt\zeta\chi^2&{\frac {\sqrt{1407}}{112}}\Phi\frac{M^{2}}{r^{2}} \sin^2(\theta),\\
 k^{t(2,1)}_{1,\mathrm{CS}}=&~\zeta\chi^2&\frac{M^{2}}{r^{2}}\Bigg[
\frac {67}{1792}\frac{1}{{f}^{2}}
\left( 
1
+{\frac {428}{603}}\frac{M}{r}
-{\frac {1651}{4221}}\frac{M^{2}}{r^{2}}
-{\frac {198860}{4221}}\frac{M^{3}}{r^{3}}
-{\frac {177110}{4221}}\frac{M^{4}}{r^{4}}
-{\frac {8984}{201}}\frac{M^{5}}{r^{5}}\right.\nonumber\\&&\left.
+{\frac {24400}{67}}\frac{M^{6}}{r^{6}}
+{\frac {10752}{67}}\frac{M^{7}}{r^{7}}
\right)\nonumber\\
&&-\frac{67}{3584}\left( 
1
-{\frac {778}{603}}\frac{M}{r}
-{\frac {54620}{4221}}\frac{M^{2}}{r^{2}}
-{\frac {131980}{1407}}\frac{M^{3}}{r^{3}}
-{\frac {24140}{67}}\frac{M^{4}}{r^{4}}\right.\nonumber\\&&\left.
-{\frac {48168}{67}}\frac{M^{5}}{r^{5}}
-{\frac {56448}{67}}\frac{M^{6}}{r^{6}}\right)\Bigg] \left( 3 \cos^2(\theta)-1 \right),\\
k^{\theta(2,\frac{1}{2})}_{1,\mathrm{CS}}=
&~\sqrt\zeta\chi^2&{\frac {28793 \sqrt{3}\sqrt{469} }{945504}\frac{1}{M\Phi} \frac{M^{3}}{r^{3}}\left( 
1
+{\frac {159962}{28793}}\frac{M}{r}
+{\frac {1774370}{86379}}\frac{M^{2}}{r^{2}}
+{\frac {3490396}{86379}}\frac{M^{3}}{r^{3}}
+{\frac {1497720}{28793}}\frac{M^{4}}{r^{4}}
\right)\cos \left( \theta \right) \sin \left( \theta \right), }\\
k^{\theta(2,1)}_{1,\mathrm{CS}}=&~\zeta\chi^2&{\frac {201}{448}\frac{1}{M}\frac{M^{3}}{r^{3}} \left( 
1
+{\frac {7489}{2412}}\frac{M}{r}
+{\frac {63395}{8442}}\frac{M^{2}}{r^{2}}
-{\frac {4955}{1407}}\frac{M^{3}}{r^{3}}
-{\frac {4505}{201}}\frac{M^{4}}{r^{4}}
-{\frac {5273}{67}}\frac{M^{5}}{r^{5}}
+{\frac {2016}{67}}\frac{M^{6}}{r^{6}}
\right)\cos \left( \theta \right) \sin \left( \theta \right), }  \\
k^{\phi(1,
\frac{1}{2})}_{1,\mathrm{CS}}=
 &~\sqrt\zeta\chi&{\frac {\sqrt{1407}}{112M}}\frac{\Phi}{M}\frac{M^{2}}{r^{2}},\\
k^{\phi(2,1)}_{1,\mathrm{CS}}=&~-\zeta\chi&{\frac {25}{16}\frac{1}{Mf}\frac{M^{5}}{r^{5}}\left(
1
+{\frac{4}{5}}\frac{M}{r}
+{\frac {162}{175}}\frac{M^{2}}{r^{2}}
-9\frac{M^{3}}{r^{3}}
\right), }
\end{eqnarray}
where
\begin{align}
  \Phi =&~   \sqrt{1 + \frac{2840}{603}\frac{M}{r} + \frac{64660}{4221}\frac{M^2}{r^2}   +\frac{1740}{67}\frac{M^3}{r^3} + \frac{1980}{67}\frac{M^4}{r^4}}.
\end{align}
\end{widetext}
The nonvanishing components of $\delta k^\alpha_{2,3,4,\mathrm{CS}}$ can then be constructed using the terms above in the following way
\begin{align}
     \delta k^t_{2,\mathrm{CS}}  =&~  - k^{t(2,\frac{1}{2})}_{1,\mathrm{CS}} +k^{t(2,1)}_{1,\mathrm{CS}},\nonumber\\
     \delta k^\theta_{2,\mathrm{CS}}  =&~  - k^{\theta(2,\frac{1}{2})}_{1,\mathrm{CS}} +k^{\theta(2,1)}_{1,\mathrm{CS}},\nonumber\\
     \delta k^\phi_{2,\mathrm{CS}}  =&~  - k^{\phi(1,\frac{1}{2})}_{1,\mathrm{CS}} +k^{\phi(1,1)}_{1,\mathrm{CS}},\nonumber\\
     \delta k^t_{3,\mathrm{CS}}  =&~   \delta k^t_{1,\mathrm{CS}},\nonumber\\
     \delta k^\theta_{3,\mathrm{CS}}  =&~  - \delta k^\theta_{1,\mathrm{CS}}\nonumber\\
     \delta k^\phi_{3,\mathrm{CS}}  =&~ \delta k^\phi_{1,\mathrm{CS}},\nonumber\\
     \delta k^t_{4,\mathrm{CS}}  =&~   \delta k^t_{2,\mathrm{CS}},\nonumber\\
     \delta k^\theta_{4,\mathrm{CS}}  =&~  - \delta k^\theta_{2,\mathrm{CS}}\nonumber\\
     \delta k^\phi_{4,\mathrm{CS}}  =&~ \delta k^\phi_{2,\mathrm{CS}}.
\end{align}

The complex null tetrad and Weyl scalars of dCS gravity in a frame where $l^\alpha$ is aligned with a PND and $\Psi_0=\Psi_4 = 0$ are 
\begin{align}
l^\alpha_\mathrm{CS}=&~l^\alpha_\mathrm{GR} +\delta l^\alpha_{\mathrm{CS}}+\mathcal{O}(\chi^3\sqrt{\zeta}),\nonumber\\
n^\alpha_\mathrm{CS}=&~n^\alpha_\mathrm{GR}+\delta n^\alpha_{\mathrm{CS}}+\mathcal{O}(\chi^3\sqrt{\zeta}),\nonumber\\
m^\alpha_\mathrm{CS}=&~m^\alpha_\mathrm{GR}+\delta m^\alpha_{\mathrm{CS}}+\mathcal{O}(\chi^3\sqrt{\zeta}),\nonumber\\
\Psi_1^\mathrm{CS} =&~  \delta \Psi_1^\mathrm{CS}+\mathcal{O}(\chi^3\sqrt{\zeta}) ,  \nonumber\\
\Psi_2^\mathrm{CS} =&~ \Psi_2^\mathrm{GR} + \delta \Psi_2^\mathrm{CS} +\mathcal{O}(\chi^3{\zeta}),  \nonumber\\
\Psi_3^\mathrm{CS} =&~  \delta \Psi_3^\mathrm{CS}+\mathcal{O}(\chi^3\sqrt{\zeta}) ,  \nonumber\\
\end{align} 
where the GR parts are given in Sec. \ref{SEC:Basics in GR} and 
\begin{align}
    \delta l^\alpha_{\mathrm{CS}} = \delta k^\alpha_{1\mathrm{CS}}.
\end{align}
The nonvanishing  components of the remaining dCS corrections are
\begin{widetext}
\begin{eqnarray}
\delta n^{t}_{\mathrm{CS}}=
&~\sqrt\zeta\chi^2&{\frac { \sqrt{1407}}{224}}\Phi f\frac{M^{2}}{r^{2}}\sin^2(\theta) \nonumber\\
&~+\zeta\chi^2&\frac{M^{2}}{r^{2}} \left[ \frac {67}{3584} \Bigg(
1
+{\frac {1634}{603}}\frac{M}{r}
+{\frac {3175}{469}}\frac{M^{2}}{r^{2}}
-{\frac {119660}{4221}}\frac{M^{3}}{r^{3}}
+{\frac {2710}{201}}\frac{M^{4}}{r^{4}}
+{\frac {3432}{67}}\frac{M^{5}}{r^{5}}
+{\frac {35448}{67}}\frac{M^{6}}{r^{6}}
\right)\nonumber\\
&&-\frac {67}{7168}\frac{M^{2}}{r^{2}} \left( 
1
+{\frac {1634}{603}}\frac{M}{r}
+{\frac {1950}{469}}\frac{M^{2}}{r^{2}}
-{\frac {166700}{4221}}\frac{M^{3}}{r^{3}}
+{\frac {9500}{201}}\frac{M^{4}}{r^{4}}
-{\frac {15048}{67}}\frac{M^{5}}{r^{5}}\right.\nonumber\\&&\left.
-{\frac {5096}{67}}\frac{M^{6}}{r^{6}}
-{\frac {129024}{67}}\frac{M^{7}}{r^{7}}
\right) \left( 3 \cos^2(\theta)-1\right)\Bigg], \\
\delta n^{r}_{\mathrm{CS}}=
&~-\zeta\chi^2&\frac{M^{3}}{r^{3}}\Bigg[\frac {25}{768} \frac{M}{r} \left( 
1
+3\frac{M}{r}
+{\frac {322}{5}}\frac{M^{2}}{r^{2}}
+{\frac {198}{5}}\frac{M^{3}}{r^{3}}
+{\frac {6276}{175}}\frac{M^{4}}{r^{4}}
-{\frac {17496}{25}}\frac{M^{5}}{r^{5}}\right)\nonumber\\
&&-\frac {201}{3584}f\left( 
1
+{\frac {1459}{603}}\frac{M}{r}
+{\frac {20000}{4221}}\frac{M^{2}}{r^{2}}
+{\frac {51580}{1407}}\frac{M^{3}}{r^{3}}
-{\frac {7580}{201}}\frac{M^{4}}{r^{4}}\right.\nonumber\\&&\left.
-{\frac {22492}{201}}\frac{M^{5}}{r^{5}}
-{\frac {40320}{67}}\frac{M^{6}}{r^{6}}
\right)\left( 3 \cos^2(\theta)-1 \right)\Bigg],\\
\delta n^{\theta}_{\mathrm{CS}}=
&~-\sqrt\zeta\chi^2&\frac {28793\sqrt{1407} }{1891008}\frac{f}{M\Phi}\frac{M^{3}}{r^{3}}  \left( 
1
+{\frac {159962}{28793}}\frac{M}{r}
+{\frac {1774370}{86379}}\frac{M^{2}}{r^{2}}
+{\frac {3490396}{86379}}\frac{M^{3}}{r^{3}}
+{\frac {1497720}{28793}}\frac{M^{4}}{r^{4}}
\right)\cos \left( \theta \right) \sin \left( \theta \right)  \nonumber\\
&~-\zeta\chi^2&\frac {201f}{896M} \frac {f}{M}\frac{M^{3}}{r^{3}}\left( 
1
+{\frac {7489}{2412}}\frac{M}{r}
+{\frac {63395}{8442}}\frac{M^{2}}{r^{2}}
-{\frac {4955}{1407}}\frac{M^{3}}{r^{3}}
-{\frac {4505}{201}}\frac{M^{4}}{r^{4}}\right.\nonumber\\&&\left.
-{\frac {5273}{67}}\frac{M^{5}}{r^{5}}
+{\frac {2016}{67}}\frac{M^{6}}{r^{6}}
\right)\cos \left( \theta \right) \sin \left( \theta \right),  \\
\delta n^{\phi}_{\mathrm{CS}}=
 &~\sqrt\zeta\chi&{\frac { \sqrt{1407}}{224}}\frac{\Phi f}{M}\frac{M^{2}}{r^{2}} -\zeta\chi{\frac{25}{32}\frac{1}{M} \frac{M^{5}}{r^{5}}\left( 
1
+\frac{4}{5}\frac{M}{r}
+{\frac {162}{175}}\frac{M^{2}}{r^{2}}
-9\frac{M^{3}}{r^{3}}
\right)},\\
\delta m^{t}_{\mathrm{CS}}=
 &~\sqrt\zeta\chi&{\frac {i\sqrt{2814}}{224}}\Phi\frac{M}{r}\sin \left( \theta \right)
+~\sqrt\zeta\chi^2{\frac {\sqrt{2814}}{224}}\Phi\frac{M^{2}}{r^{2}} 
\cos \left( \theta \right) \sin \left( \theta \right), \\
\delta m^{r}_{\mathrm{CS}}=
&~-\sqrt\zeta\chi^2&{\frac {28793 \sqrt{2814} }{1891008}\frac{f}{\Phi} \frac{M^{2}}{r^{2}} \left( 
1
+{\frac {159962}{28793}}\frac{M}{r}
+{\frac {1774370}{86379}}\frac{M^{2}}{r^{2}}
+{\frac {3490396}{86379}}\frac{M^{3}}{r^{3}}
+{\frac {1497720}{28793}}\frac{M^{4}}{r^{4}}
\right) \cos \left( \theta \right)\sin \left( \theta \right) } \nonumber\\
&~-\zeta\chi^2&\frac {201 \sqrt{2} }{896} f\frac{M^{2}}{r^{2}}\left( 
1
+{\frac {7489}{2412}}\frac{M}{r}
+{\frac {63395}{8442}}\frac{M^{2}}{r^{2}}
-{\frac {4955}{1407}}\frac{M^{3}}{r^{3}}
-{\frac {4505}{201}}\frac{M^{4}}{r^{4}}
\right.\nonumber\\&&\left.-{\frac {5273}{67}}\frac{M^{5}}{r^{5}}+
{\frac {2016}{67}}\frac{M^{6}}{r^{6}}
\right)\sin \left( \theta \right)\cos \left( \theta \right) ,\\
\delta m^{\theta}_{\mathrm{CS}}=
&~\sqrt\zeta\chi^2&{{\frac {i28793}{1891008 \sqrt{2814}}} \frac{\sqrt{f}}{M\Phi} \frac{M^{3}}{r^{3}}\left(
1
+{\frac {159962}{28793}}\frac{M}{r}
+{\frac {1774370}{86379}}\frac{M^{2}}{r^{2}}
+{\frac {3490396}{86379}}\frac{M^{3}}{r^{3}}
+{\frac {1497720}{28793}}\frac{M^{4}}{r^{4}}
\right) \cos \left( \theta \right) \sin \left( \theta \right)} \nonumber\\
&~+\zeta\chi^2&\Bigg[{{\frac{i201\sqrt{2}}{896}} \frac {\sqrt{f} }{M} \frac{M^{3}}{r^{3}}\left( 
1
+{\frac {7489}{2412}}\frac{M}{r}
+{\frac {63395}{8442}}\frac{M^{2}}{r^{2}}
-{\frac {4955}{1407}}\frac{M^{3}}{r^{3}}
-{\frac {4505}{201}}\frac{M^{4}}{r^{4}}
-{\frac {5273}{67}}\frac{M^{5}}{r^{5}}
+{\frac {2016}{67}}\frac{M^{6}}{r^{6}}
\right) \cos \left( \theta \right) \sin \left( \theta \right)}\nonumber\\
&&-\frac {201 \sqrt{2}}{7168}\frac{1}{M} \frac{M^{4}}{r^{4}}\left( 
1
+{\frac {1420}{603}}\frac{M}{r}
+{\frac {18908}{4221}}\frac{M^{2}}{r^{2}}
+{\frac {1480}{603}}\frac{M^{3}}{r^{3}}
+{\frac {22460}{1407}}\frac{M^{4}}{r^{4}}\right.\nonumber\\&&\left.
+{\frac {3848}{201}}\frac{M^{5}}{r^{5}}
+{\frac {5376}{67}}\frac{M^{6}}{r^{6}}
\right)\left( 3 \cos^2(\theta)-1 \right) \Bigg],\\
\delta m^{\phi}_{\mathrm{CS}}=
&~\sqrt\zeta\chi^2&{{\frac {i28793\sqrt{2814}}{1891008}} \frac{\sqrt{f}}{M\Phi} \frac{M^{3}}{r^{3}}\left( 
1
+{\frac {159962}{28793}}\frac{M}{r}
+{\frac {1774370}{86379}}\frac{M^{2}}{r^{2}}
+{\frac {3490396}{86379}}\frac{M^{3}}{r^{3}}
+{\frac {1497720}{28793}}\frac{M^{4}}{r^{4}}
\right)\cos \left( \theta \right) \sin \left( \theta \right)}\nonumber\\
&+~\zeta\chi^2&\frac{1}{M}\frac{M^{3}}{r^{3}} \Bigg[{{\frac {i67\sqrt{2}}{3584}} \left( 
1
+{\frac {1634}{603}}\frac{M}{r}
+{\frac {8300}{1407}}\frac{M^{2}}{r^{2}}
-{\frac {137300}{4221}}\frac{M^{3}}{r^{3}}
-{\frac {20180}{201}}\frac{M^{4}}{r^{4}}
-{\frac {14648}{67}}\frac{M^{5}}{r^{5}}
-{\frac {2016}{67}}\frac{M^{6}}{r^{6}}
\right)\frac{1}{\sin\left(\theta\right)} }\nonumber\\
&&-{\frac {201 \sqrt{2} }{896}\sqrt{f}\left( 
1
+{\frac {7489}{2412}}
+{\frac {63395}{8442}}\frac{M^{2}}{r^{2}}
-{\frac {4955}{1407}}\frac{M^{3}}{r^{3}}
-{\frac {4505}{201}}\frac{M^{4}}{r^{4}}
-{\frac {5273}{67}}\frac{M^{5}}{r^{5}}
+{\frac {2016}{67}}\frac{M^{6}}{r^{6}}
\right)\cos \left( \theta \right) }\nonumber\\
&&-{\frac {i67\sqrt{2}}{7168}}   \left( 
1
+{\frac {3443}{603}}\frac{M}{r}
+{\frac {6080}{469}}\frac{M^{2}}{r^{2}}
-{\frac {80576}{4221}}\frac{M^{3}}{r^{3}}
-{\frac {18700}{201}}\frac{M^{4}}{r^{4}}
-{\frac {80076}{469}}\frac{M^{5}}{r^{5}}\right.\nonumber\\&&\left.
+{\frac {1832}{67}}\frac{M^{6}}{r^{6}}
+{\frac {16128}{67}}\frac{M^{7}}{r^{7}}
\right)\frac {\left( 3 \cos^2(\theta)-1 \right) }{\sin \left( \theta \right)}  \Bigg],\\
\delta \Psi_1^{\mathrm{CS}}=
 &~-\sqrt\zeta\chi&{{\frac {i3\sqrt{2814}}{224}}\frac{\Phi}{{{M}^{2}}}\frac{M^{4}}{r^{4}} \sin \left( \theta \right) }\nonumber\\
&~-\sqrt\zeta\chi^2&\frac {11909\sqrt{2814} }{630336}\frac{1}{{M}^{2}\Phi} \frac{M^{5}}{r^{5}} \left( 
1
+{\frac {80442}{11909}} \frac{M}{r}
+{\frac {998450}{35727}}\frac{M^{2}}{r^{2}}\right.\nonumber\\&&\left.
+{\frac {2174956}{35727}}\frac{M^{3}}{r^{3}}
+{\frac {998760}{11909}}\frac{M^{4}}{r^{4}}
\right)\cos \left( \theta \right) \sin \left( \theta \right) ,\\
\delta \Psi_2^{\mathrm{CS}}=
&~\zeta\chi&{{\frac {i15}{8}}\frac{1}{{M}^{2}}\frac{M^{7}}{r^{7}} \left( 
1
+2\frac{M}{r}
+{\frac {18}{5}}\frac{M^{2}}{r^{2}}
\right)\cos \left( \theta \right) } \nonumber\\
&~-\zeta\chi^2&\frac{1}{{M}^{2}} \frac{M^{5}}{r^{5}}\Bigg[{\frac {201}{1792} \left( 
1
+{\frac {1459}{603}}\frac{M}{r}
+{\frac {7075}{1407}}\frac{M^{2}}{r^{2}}
-{\frac {577520}{12663}}\frac{M^{3}}{r^{3}}
-{\frac {5490}{67}}\frac{M^{4}}{r^{4}}
-{\frac {10052}{67}}\frac{M^{5}}{r^{5}}
+{\frac {49896}{67}}\frac{M^{6}}{r^{6}}
\right) }\nonumber\\
&&+\frac {1005 }{3584} \left( 
1
+{\frac {3871}{3015}}\frac{M}{r}
+{\frac {1066}{603}}\frac{M^{2}}{r^{2}}
-{\frac {163964}{12663}}\frac{M^{3}}{r^{3}}
-{\frac {2500}{201}}\frac{M^{4}}{r^{4}}
-{\frac {2004}{67}}\frac{M^{5}}{r^{5}}\right.\nonumber\\&&\left.
+{\frac {101992}{335}}\frac{M^{6}}{r^{6}}
-{\frac {16128}{335}}\frac{M^{7}}{r^{7}}
\right)\left( 3 \cos^2(\theta)-1 \right)\Bigg]\\
\delta \Psi_3^{\mathrm{CS}}=
 &~\sqrt\zeta\chi&{\frac {i3\sqrt{2814}}{448}} \frac {\Phi f}{{M}^{2}}\frac{M^{4}}{r^{4}}\sin \left( \theta \right)\nonumber\\
&~-\sqrt\zeta\chi^2&{\frac {4975 \sqrt{2814} }{1260672}\frac {f }{{M}^{2}\Phi}\frac{M^{5}}{r^{5}} \left( 
1
-{\frac {922}{4975}}\frac{M}{r}
-{\frac {44506}{2985}}\frac{M^{2}}{r^{2}}
-{\frac {859516}{14925}}\frac{M^{3}}{r^{3}}
-{\frac {19992}{199}}\frac{M^{4}}{r^{4}}
\right)\cos \left( \theta \right) \sin \left( \theta \right) }
\end{eqnarray}
\end{widetext}

\subsection{Scalar Gauss-Bonnet Gravity}
Here we extend the results presented in Section \ref{SEC:sGBPNDs}, including sGB corrections through $\mathcal{O}(\chi^2\zeta)$ to the PND, tetrad and Weyl scalars. The full results, which include corrections to $\mathcal{O}(\chi^4\zeta),$ are collected in a \textit{Mathematica} notebook that is provided in the Supplemental Material.

The PNDs of sGB are
\begin{align}
k^\alpha_{1,\mathrm{GB}}=&~k^\alpha_{1,\mathrm{GR}} + \delta k^\alpha_{1,\mathrm{GB}} +\mathcal{O}(\chi^3\sqrt{\zeta}), \nonumber\\
k^\alpha_{2,\mathrm{GB}}=&~k^\alpha_{1,\mathrm{GR}} + \delta k^\alpha_{2,\mathrm{GB}}+\mathcal{O}(\chi^3\sqrt{\zeta}), \nonumber\\
k^\alpha_{3,\mathrm{GB}}=&~k^\alpha_{2,\mathrm{GR}} + \delta k^\alpha_{3,\mathrm{GB}}+\mathcal{O}(\chi^3\sqrt{\zeta}), \nonumber\\
k^\alpha_{4,\mathrm{GB}}=&~k^\alpha_{2,\mathrm{GR}} + \delta k^\alpha_{4,\mathrm{GB}}+\mathcal{O}(\chi^3\sqrt{\zeta}),\nonumber\\
\end{align}
where the $k^\alpha_{1,2,\mathrm{GR}}$ are given in Section \ref{SEC:Basics in GR} and 
\begin{align}\delta k^\alpha_{i,\mathrm{GB}} =
 k^{\alpha(1,\frac{1}{2})}_{i,\mathrm{GB}} 
+  k^{\alpha(2,\frac{1}{2})}_{i,\mathrm{GB}} 
+ k^{\alpha(0,1)}_{i,\mathrm{GB}} 
+ k^{\alpha(1,1)}_{i,\mathrm{GB}} 
+ k^{\alpha(2,1)}_{i,\mathrm{GB}},
\end{align}
with  $ k^{\alpha(m,n)}_{i,\mathrm{GB}}\propto \chi^m\zeta^n$. The terms that make up the nonvanishing components of $\delta k^\alpha_{1\mathrm{GB}}$ are
\begin{widetext}
\begin{eqnarray}
k^{t(0,1)}_{1,\mathrm{GB}}=
 &~-\zeta&\frac{1}{2}\frac{1}{{f}^{2}}\frac{M^{2}}{r^{2}} \left( 
1
+\frac{4}{3}\frac{M}{r}
+26\frac{M^{2}}{r^{2}}
+{\frac{32}{5}}\frac{M^{3}}{r^{3}}
+{\frac{48}{5}}\frac{M^{4}}{r^{4}}
-{\frac{448}{3}}\frac{M^{5}}{r^{5}} \right),\\
k^{t(2,1)}_{1,\mathrm{GB}}=&~-\zeta\chi^2&\Bigg[\frac{25727}{31500}\frac{1}{{f}^{3}} \frac{M^{2}}{r^{2}}\left( 1
-{\frac{19602}{25727}}\frac{M}{r}
+{\frac{435738}{180089}}\frac{M^{2}}{r^{2}}
-{\frac{2325412}{180089}}\frac{M^{3}}{r^{3}}
+{\frac{3839020}{180089}}\frac{M^{4}}{r^{4}}
+{\frac{9856880}{180089}}\frac{M^{5}}{r^{5}}\right.\nonumber\\&&\left.
+{\frac{225800}{1979}}\frac{M^{6}}{r^{6}}
-{\frac{15916400}{25727}}\frac{M^{7}}{r^{7}}
+{\frac{10539200}{25727}}\frac{M^{8}}{r^{8}}
+{\frac{3920000}{25727}}\frac{M^{9}}{r^{9}} \right) \nonumber\\
&&-\frac{4463}{15750}\frac{M^{2}}{r^{2}} \left( 
1
+{\frac{49968}{4463}}\frac{M}{r}
+{\frac{1433940}{31241}}\frac{M^{2}}{r^{2}}
+{\frac{4037610}{31241}}\frac{M^{3}}{r^{3}}
+{\frac{634650}{4463}}\frac{M^{4}}{r^{4}}
+{\frac{676250}{4463}}\frac{M^{5}}{r^{5}}\right.\nonumber\\&&\left.
-{\frac{940800}{4463}}\frac{M^{6}}{r^{6}}
+{\frac{2520000}{4463}}\frac{M^{7}}{r^{7}} \right) \left( 3 \cos^{2}(\theta) -1 \right)\Bigg],\\
k^{\theta(1,
\frac{1}{2})}_{1,\mathrm{GB}}=
 &~\sqrt\zeta\chi&\frac{\sqrt{468615} }{525}\frac{\Theta}{M}\frac{M^{2}}{r^{2}}\sin \left( \theta \right),\\
 k^{\theta(2,1)}_{1,\mathrm{GB}}=&~-\zeta\chi^2&\frac{17852 }{2625}\frac{1}{M}\frac{M^{3}}{r^{3}} \left( 
1
+{\frac{13236}{4463}}\frac{M}{r}
+{\frac{445695}{62482}}\frac{M^{2}}{r^{2}}
-{\frac{382545}{249928}}\frac{M^{3}}{r^{3}}
-{\frac{253475}{8926}}\frac{M^{4}}{r^{4}}
-{\frac{1660625}{17852}}\frac{M^{5}}{r^{5}}\right.\nonumber\\&&\left.
-{\frac{33600}{4463}}\frac{M^{6}}{r^{6}}
+{\frac{78750}{4463}}\frac{M^{7}}{r^{7}} \right)\cos \left( \theta \right)\sin \left( \theta \right), \\
k^{\phi(2,
\frac{1}{2})}_{1,\mathrm{GB}}=
 &~-\sqrt\zeta\chi^2&\frac{70797 \sqrt{468615}}{10934350}\frac{1}{M\Theta} \frac{M^{3}}{r^{3}}\left( 
1
+{\frac{132270}{23599}}\frac{M}{r}
+{\frac{13347430}{637173}}\frac{M^{2}}{r^{2}}
+{\frac{26484500}{637173}}\frac{M^{3}}{r^{3}}\right.\nonumber\\&&\left.
+{\frac{11384800}{212391}} \frac{M^{4}}{r^{4}}\right) \cos \left( \theta \right),\\
k^{\phi(1,
1)}_{1,\mathrm{GB}}=
 &~\sqrt\zeta\chi^2&\frac{1}{3}{\frac{1}{M{f}^{2}} \frac{M^{3}}{r^{3}}\left( 
1
-\frac{7}{2}\frac{M}{r}
+{\frac{46}{5}}\frac{M^{2}}{r^{2}}
-{\frac{337}{5}}\frac{M^{3}}{r^{3}}
-{\frac{36}{5}}\frac{M^{4}}{r^{4}}
-{\frac{536}{5}}\frac{M^{5}}{r^{5}}
+448\frac{M^{6}}{r^{6}} \right), }
\end{eqnarray}
where
\begin{align}
\Theta =&~ \sqrt{1+\frac{21440}{4463}\frac{M}{r}+\frac{508960}{31241}\frac{M^2}{r^2}+\frac{135300}{4463}\frac{M^3}{r^3}+\frac{167600}{4463}\frac{M^4}{r^4}}.
\end{align}
\end{widetext}
The nonvanishing components of $\delta k^\alpha_{2,3,4,\mathrm{GB}}$ can then be constructed using the terms above in the following way
\begin{align}
     \delta k^t_{2,\mathrm{GB}}  =&~  \delta k^t_{1,\mathrm{GB}},\nonumber\\
     \delta k^\theta_{2,\mathrm{GB}}  =&~  - k^{\theta(1,\frac{1}{2})}_{1,\mathrm{GB}} +k^{\theta(2,1)}_{1,\mathrm{GB}},\nonumber\\
     \delta k^\phi_{2,\mathrm{GB}}  =&~  - k^{\phi(2,\frac{1}{2})}_{1,\mathrm{GB}} +k^{\phi(1,1)}_{1,\mathrm{GB}},\nonumber\\
     \delta k^t_{3,\mathrm{GB}}  =&~   -\delta k^t_{1,\mathrm{GB}},\nonumber\\
     \delta k^\theta_{3,\mathrm{GB}}  =&~  \delta k^\theta_{1,\mathrm{GB}}\nonumber\\
     \delta k^\phi_{3,\mathrm{GB}}  =&~ -\delta k^\phi_{1,\mathrm{GB}},\nonumber\\
     \delta k^t_{4,\mathrm{GB}}  =&~   -\delta k^t_{1,\mathrm{GB}},\nonumber\\
     \delta k^\theta_{4,\mathrm{GB}}  =&~  \delta k^\theta_{2,\mathrm{GB}}\nonumber\\
     \delta k^\phi_{4,\mathrm{GB}}  =&~ -\delta k^\phi_{2,\mathrm{GB}}.
\end{align}

The complex null tetrad and Weyl scalars of sGB gravity in a frame where $l^\alpha$ is aligned with a PND and $\Psi_0=\Psi_4 = 0$ are 
\begin{align}
l^\alpha_\mathrm{GB}=&~l^\alpha_\mathrm{GR} +\delta l^\alpha_{\mathrm{GB}} +\mathcal{O}(\chi^3\sqrt{\zeta}),\nonumber\\
n^\alpha_\mathrm{GB}=&~n^\alpha_\mathrm{GR}+\delta n^\alpha_{\mathrm{GB}}+\mathcal{O}(\chi^3\sqrt{\zeta}),\nonumber\\
m^\alpha_\mathrm{GB}=&~m^\alpha_\mathrm{GR}+\delta m^\alpha_{\mathrm{GB}}+\mathcal{O}(\chi^3\sqrt{\zeta}),\nonumber\\
\Psi_1^\mathrm{GB} =&~  \delta \Psi_1^\mathrm{GB}+\mathcal{O}(\chi^3\sqrt{\zeta}) ,  \nonumber\\
\Psi_2^\mathrm{GB} =&~ \Psi_2^\mathrm{GR} + \delta \Psi_2^\mathrm{GB} +\mathcal{O}(\chi^3{\zeta}),  \nonumber\\
\Psi_3^\mathrm{GB} =&~  \delta \Psi_3^\mathrm{GB} +\mathcal{O}(\chi^3\sqrt{\zeta}),  \nonumber\\
\end{align} 
where the GR parts are given in section \ref{SEC:Basics in GR} and 
\begin{align}
    \delta l^\alpha_{\mathrm{GB}} = \delta k^\alpha_{1\mathrm{GB}}.
\end{align}
The nonvanishing  components of the remaining dCS corrections are
\begin{widetext}
\begin{eqnarray}
\delta n^{t}_{\mathrm{GB}}=
&~+\zeta&\frac{1}{4}\frac{M^{2}}{r^{2}} \left( 
1
+\frac{8}{3}\frac{M}{r}
+14\frac{M^{2}}{r^{2}}
+{\frac{128}{5}}\frac{M^{3}}{r^{3}}
+48\frac{M^{4}}{r^{4}} \right)\nonumber\\
&~-\zeta\chi^2&\frac{M^{2}}{r^{2}}\Bigg[\frac{25727}{63000} \left( 
1
+{\frac{64056}{25727}}\frac{M}{r}
+{\frac{106650}{13853}}\frac{M^{2}}{r^{2}}
+{\frac{334120}{180089}}\frac{M^{3}}{r^{3}}
+{\frac{29000}{25727}}\frac{M^{4}}{r^{4}}
+{\frac{99800}{25727}}\frac{M^{5}}{r^{5}}\right.\nonumber\\&&\left.
+{\frac{4093600}{25727}}\frac{M^{6}}{r^{6}}
+{\frac{1120000}{25727}}\frac{M^{7}}{r^{7}} \right) \nonumber\\
&&-\frac{4463}{31500}   \left( 
1
+{\frac{10764}{4463}}\frac{M}{r}
+{\frac{153675}{31241}}\frac{M^{2}}{r^{2}}
-{\frac{475070}{31241}}\frac{M^{3}}{r^{3}}
-{\frac{639725}{4463}}\frac{M^{4}}{r^{4}}
-{\frac{296350}{4463}}\frac{M^{5}}{r^{5}}\right.\nonumber\\&&\left.
-{\frac{834050}{4463}}\frac{M^{6}}{r^{6}}
+{\frac{4670400}{4463}}\frac{M^{7}}{r^{7}}
-{\frac{5670000}{4463}} \frac{M^{8}}{r^{8}}\right)\left( 3 \cos^{2}(\theta) -1 \right)\Bigg], \\
\delta n^{r}_{\mathrm{GB}}=
&~-\zeta&\frac{1}{2}\frac{M^{2}}{r^{2}} \left( 
1
+
\frac{M}{r}
+{\frac{52}{3}}\frac{M^{2}}{r^{2}}
+2\frac{M^{3}}{r^{3}}
+{\frac{16}{5}}\frac{M^{4}}{r^{4}}
-{\frac{368}{3}}\frac{M^{5}}{r^{5}} \right)\nonumber\\
&~+\zeta\chi^2&\frac{M^{2}}{r^{2}}\Bigg[\frac{25727}{31500} \left( 
1
+{\frac{22227}{25727}}\frac{M}{r}
+{\frac{648916}{180089}}\frac{M^{2}}{r^{2}}
-{\frac{1880180}{180089}}\frac{M^{3}}{r^{3}}
-{\frac{66130}{13853}}\frac{M^{4}}{r^{4}}
-{\frac{13100}{1979}}\frac{M^{5}}{r^{5}}
+{\frac{2885300}{25727}}\frac{M^{6}}{r^{6}}\right.\nonumber\\&&\left.
-{\frac{8982400}{25727}}\frac{M^{7}}{r^{7}}
-{\frac{2100000}{25727}}\frac{M^{8}}{r^{8}}\right)\nonumber\\
&&-\frac{4463  }{15750}\left( 
1
+{\frac{16977}{4463}}\frac{M}{r}
+{\frac{72202}{31241}}\frac{M^{2}}{r^{2}}
-{\frac{552983}{31241}}\frac{M^{3}}{r^{3}}
-{\frac{9977815}{62482}}\frac{M^{4}}{r^{4}}
+{\frac{4939045}{31241}}\frac{M^{5}}{r^{5}}
-{\frac{686875}{4463}}\frac{M^{6}}{r^{6}}\right.\nonumber\\&&\left.
+{\frac{7859350}{4463}}\frac{M^{7}}{r^{7}}
-{\frac{14427000}{4463}}\frac{M^{8}}{r^{8}}
+{\frac{10710000}{4463}}\frac{M^{9}}{r^{9}} \right)\left( 3 \cos^{2}(\theta) -1 \right),\\
\delta n^{\theta}_{\mathrm{GB}}=
&~-\sqrt\zeta\chi&{\frac{ \sqrt{468615} }{1050M}}\Phi f\frac{M^{2}}{r^{2}}\sin \left( \theta \right)\nonumber\\
&~+\zeta\chi^2&\frac{8926 }{2625}
\frac{1}{M}\frac{M^{3}}{r^{3}}\left( 
1
+{\frac{4310}{4463}}\frac{M}{r}
+{\frac{75087}{62482}}\frac{M^{2}}{r^{2}}
-{\frac{564015}{35704}}\frac{M^{3}}{r^{3}}
-{\frac{3166105}{124964}}\frac{M^{4}}{r^{4}}
-{\frac{646725}{17852}}\frac{M^{5}}{r^{5}}
+{\frac{1593425}{8926}}\frac{M^{6}}{r^{6}}\right.\nonumber\\&&\left.
+{\frac{145950}{4463}}\frac{M^{7}}{r^{7}}
-{\frac{157500}{4463}}\frac{M^{8}}{r^{8}}\right)\cos \left( \theta \right) \sin \left( \theta \right), \\
\delta n^{\phi}_{\mathrm{GB}}=
&~-\sqrt\zeta\chi^2&{\frac{70797 \sqrt{468615}}{21868700}\frac{f}{M\Theta}\frac{M^{3}}{r^{3}} \left( 
1
+{\frac{132270}{23599}}\frac{M}{r}
+{\frac{13347430}{637173}}\frac{M^{2}}{r^{2}}
+{\frac{26484500}{637173}}\frac{M^{3}}{r^{3}}
+{\frac{11384800}{212391}}\frac{M^{4}}{r^{4}} \right) \cos \left( \theta \right)}\nonumber\\
&~+\zeta\chi&\frac{1}{6}{\frac{1}{M}\frac{M^{3}}{r^{3}} \left( 
1
+\frac{3}{2}\frac{M}{r}
+{\frac{76}{5}}\frac{M^{2}}{r^{2}}
+15\frac{M^{3}}{r^{3}}
+{\frac{144}{5}}\frac{M^{4}}{r^{4}}
-40\frac{M^{5}}{r^{5}} \right) },\\
\delta m^{t}_{\mathrm{GB}}=
&~-\sqrt\zeta\chi^2&\frac{M^{2}}{r^{2}}\Bigg[{\frac{\sqrt{937230}}{1050}}\Theta \sqrt{f}  \sin^{2}(\theta)\nonumber\\
&&+{\frac{i70797\sqrt{937230}}{21868700}}\frac{1 }{\Theta} \left(
1
+{\frac{132270}{23599}}\frac{M}{r}
+{\frac{13347430}{637173}}\frac{M^{2}}{r^{2}}\right.\nonumber\\&&\left.
+{\frac{26484500}{637173}}\frac{M^{3}}{r^{3}}
+{\frac{11384800}{212391}} \frac{M^{4}}{r^{4}}\right)\sin \left( \theta \right)  \cos \left( \theta \right)\Bigg]\nonumber\\
&~+\zeta\chi& \frac{i\sqrt{2}}{6} \frac{M^{2}}{r^{2}}\left(
1
+2\frac{M}{r} 
+16\frac{M^{2}}{r^{2}}
+28\frac{M^{3}}{r^{3}}
+{\frac{256}{5}}\frac{M^{4}}{r^{4}}
\right)\sin \left( \theta \right) \nonumber\\
&~+\zeta\chi^2&\frac{\sqrt{2}}{6}  \frac{M^{3}}{r^{3}} \left(1
+2\frac{M}{r} 
+16\frac{M^{2}}{r^{2}}
+28\frac{M^{3}}{r^{3}}
+{\frac{256}{5}}\frac{M^{4}}{r^{4}} \right)\cos \left( \theta \right) \sin \left( \theta \right),\\
\delta m^{r}_{\mathrm{GB}}=
&~-\sqrt\zeta\chi&\frac{ \sqrt{937230}}{1050}\Theta f\frac{M}{r}\sin \left( \theta \right)~
+\sqrt\zeta\chi^2{\frac{i\sqrt{937230}}{1050}}\Theta f \frac{M^{2}}{r^{2}} \cos \left( \theta \right)\sin \left( \theta \right)\nonumber\\
&~-\zeta\chi^2& f\frac{M^{2}}{r^{2}}\Bigg[{\frac{i4463\sqrt{2}}{5250}}{\Theta}^{2}\sqrt{f}\sin^{2}(\theta)\nonumber\\
&&-\frac{8926 \sqrt{2}}{2625}\left( 
1
+{\frac{13236}{4463}}\frac{M}{r}
+{\frac{445695}{62482}}\frac{M^{2}}{r^{2}}
-{\frac{382545}{249928}}\frac{M^{3}}{r^{3}}
-{\frac{253475}{8926}}\frac{M^{4}}{r^{4}}
-{\frac{1660625}{17852}}\frac{M^{5}}{r^{5}}\right.\nonumber\\&&\left.
-{\frac{33600}{4463}}\frac{M^{6}}{r^{6}}
+{\frac{78750}{4463}} \frac{M^{7}}{r^{7}}\right)\sin \left( \theta \right)\cos \left( \theta \right) \Bigg],\\
\delta m^{\theta}_{\mathrm{GB}}=
&~\sqrt\zeta\chi&{\frac{i\sqrt{937230}}{1050}}{\frac{\Theta\sqrt{f}}{M}}\frac{M^{2}}{r^{2}}\sin \left( \theta \right)~+\sqrt\zeta\chi^2{\frac{ \sqrt{937230}}{1050}}\frac{\Theta \sqrt{f}}{M}\frac{M^{3}}{r^{3}}\cos \left( \theta \right)\sin \left( \theta \right)\nonumber\\
&~-\zeta\chi^2&\nonumber\Bigg[{\frac{4463 \sqrt{2}}{7875}}\frac{{\Theta}^{2}f}{M}\frac{M^{3}}{r^{3}}\\
&&+{\frac{i8926\sqrt{2}}{2625}} \frac{\sqrt{f}}{M} \frac{M^{3}}{r^{3}}\left( 
1
+{\frac{13236}{4463}}\frac{M}{r}
+{\frac{445695}{62482}}\frac{M^{2}}{r^{2}}
-{\frac{382545}{249928}}\frac{M^{3}}{r^{3}}
-{\frac{253475}{8926}}\frac{M^{4}}{r^{4}}
-{\frac{1660625}{17852}}\frac{M^{5}}{r^{5}}\right.\nonumber\\&&\left.
-{\frac{33600}{4463}}\frac{M^{6}}{r^{6}}
+{\frac{78750}{4463}}\frac{M^{7}}{r^{7}} \right)\sin \left( \theta \right)\cos \left( \theta \right) \nonumber\\
&&-\frac{4463 \sqrt{2}}{15750}\frac{1}{M}\frac{M^{3}}{r^{3}} \left( 1
+{\frac{38417}{8926}}\frac{M}{r}
+{\frac{317685}{31241}}\frac{M^{2}}{r^{2}}
+{\frac{517453}{124964}}\frac{M^{3}}{r^{3}}
-{\frac{142635}{8926}}\frac{M^{4}}{r^{4}}
-{\frac{5620625}{62482}}\frac{M^{5}}{r^{5}}\right.\nonumber\\&&\left.
-{\frac{122025}{4463}}\frac{M^{6}}{r^{6}}
-{\frac{221900}{4463}}\frac{M^{7}}{r^{7}}
+{\frac{315000}{4463}}\frac{M^{8}}{r^{8}}
\right)\left( 3 \cos^{2}(\theta) -1 \right) \Bigg],\\
\delta m^{\phi}_{\mathrm{GB}}=
&~-\sqrt\zeta\chi&{\frac{\sqrt{937230}}{1050}}\frac{\Theta \sqrt{f} }{M}\frac{M^{2}}{r^{2}}+\sqrt\zeta\chi^2{\frac{i \sqrt{937230}}{1050}}{\frac{\Phi \sqrt{f} }{M}}\frac{M^{3}}{r^{3}}\cos \left( \theta \right)
\nonumber\\&&-~\zeta\chi^2\nonumber\frac{1}{M}\frac{M^{3}}{r^{3}}\Bigg[{\frac{i4463\sqrt{2}}{15750}} \left(
1 
+{\frac{10764}{4463}}\frac{M}{r}
+{\frac{184300}{31241}}\frac{M^{2}}{r^{2}}
-{\frac{250895}{31241}}\frac{M^{3}}{r^{3}}
-{\frac{125750}{4463}}\frac{M^{4}}{r^{4}}\right.\nonumber\\&&\left.
-{\frac{404850}{4463}}\frac{M^{5}}{r^{5}}
+{\frac{33600}{4463}}\frac{M^{6}}{r^{6}}
-{\frac{70000}{4463}} \frac{M^{7}}{r^{7}}
\right)\frac{1}{\sin \left( \theta \right)} \nonumber\\
&&-\frac{8926 \sqrt{2}}{2625}\sqrt{f}\left( 
1
+{\frac{13236}{4463}}\frac{M}{r}
+{\frac{445695}{62482}}\frac{M^{2}}{r^{2}}
-{\frac{382545}{249928}}\frac{M^{3}}{r^{3}}
-{\frac{253475}{8926}}\frac{M^{4}}{r^{4}}\right.\nonumber\\&&\left.
-{\frac{1660625}{17852}}\frac{M^{5}}{r^{5}}
-{\frac{33600}{4463}}\frac{M^{6}}{r^{6}}
+{\frac{78750}{4463}}\frac{M^{7}}{r^{7}} \right)\cos \left( \theta \right)\nonumber\\
&&-\frac{i4463\sqrt{2} }{31500}  \left(
1
+{\frac{24153}{4463}}\frac{M}{r}
+{\frac{402070}{31241}}\frac{M^{2}}{r^{2}}
+{\frac{298943}{62482}}\frac{M^{3}}{r^{3}}
-{\frac{62385}{4463}}\frac{M^{4}}{r^{4}}
-{\frac{3761775}{31241}}\frac{M^{5}}{r^{5}}\right.\nonumber\\&&\left.
-{\frac{210450}{4463}}\frac{M^{6}}{r^{6}}
-{\frac{513800}{4463}}\frac{M^{7}}{r^{7}}
+{\frac{630000}{4463}}\frac{M^{8}}{r^{8}}
\right)\frac{\left( 3 \cos^{2}(\theta) -1 \right)}{\sin(\theta)}\Bigg],\\
\delta \Psi_1^{\mathrm{GB}}=
&~-\sqrt\zeta\chi&{\frac{  \sqrt{937230}}{350{M}^{2}}}\Theta\frac{M^{4}}{r^{4}}\sin \left( \theta \right)\nonumber\\
&~+\sqrt\zeta\chi^2&{\frac  {{i87427}\sqrt{937230}}{21868700}}{\frac{ 1 }{{M}^{2}\Theta} \frac{M^{5}}{r^{5}}\left( 
1
+{\frac{590110}{87427}}\frac{M}{r}
+{\frac{7239910}{262281}}\frac{M^{2}}{r^{2}}
+{\frac{15119300}{262281}}\frac{M^{3}}{r^{3}}
+{\frac{6692000}{87427}}\frac{M^{4}}{r^{4}}
\right)\sin \left( \theta \right)\cos \left( \theta \right) }\nonumber\\
&~-\zeta\chi^2&{\frac{i4463 \sqrt{2}}{1750}} \frac{ \sqrt{f}}{{M}^{2}}
\frac{M^{5}}{r^{5}}\left( 
1
+{\frac{21440}{4463}} \frac{M}{r}
+{\frac{508960}{31241}}\frac{M^{2}}{r^{2}}
+{\frac{135300}{4463}}\frac{M^{3}}{r^{3}}
+{\frac{167600}{4463}}\frac{M^{4}}{r^{4}}
\right)\sin^{2}(\theta),\\
\delta \Psi_2^{\mathrm{GB}}=
&\zeta&\frac{1}{3}{\frac{1}{{M}^{2}} \frac{M^{4}}{r^{4}}\left( 
1
+\frac{1}{2}\frac{M}{r}
+72\frac{M^{2}}{r^{2}}
+7\frac{M^{3}}{r^{3}}
+{\frac{64}{5}}\frac{M^{4}}{r^{4}}
-840\frac{M^{5}}{r^{5}} \right) }~+\zeta\chi{ {24i\frac{f}{{M}^{2}}\frac{M^{7}}{r^{7}} \left(
1 
+2\frac{M}{r}
+4\frac{M^{2}}{r^{2}}
\right) \cos \left( \theta \right)}}\nonumber\\
&~-\zeta\chi^2&\Bigg[\frac{1}{6}\frac{1}{{M}^{2}}\frac{M^{4}}{r^{4}} \left( 
1
-{\frac{16977}{1750}}\frac{M}{r}
+{\frac{21166}{2625}}\frac{M^{2}}{r^{2}}
-{\frac{46927}{735}}\frac{M^{3}}{r^{3}}
+{\frac{1502309}{3675}}\frac{M^{4}}{r^{4}}
+{\frac{6260}{21}}\frac{M^{5}}{r^{5}}\right.\nonumber\\&&\left.
+{\frac{98438}{105}}\frac{M^{6}}{r^{6}}
-{\frac{187376}{15}}\frac{M^{7}}{r^{7}}
+80\frac{M^{8}}{r^{8}} \right) \nonumber\\
&&-\frac{4463  }{1050}\frac{1}{{M}^{2}} \frac{M^{5}}{r^{5}}\left( 
1+
{\frac{77273}{66945}}\frac{M}{r}
+{\frac{17048}{13389}}\frac{M^{2}}{r^{2}}
-{\frac{5475655}{187446}}\frac{M^{3}}{r^{3}}
-{\frac{545189}{13389}}\frac{M^{4}}{r^{4}}
-{\frac{963955}{13389}}\frac{M^{5}}{r^{5}}\right.\nonumber\\&&\left.
+{\frac{6644110}{13389}}\frac{M^{6}}{r^{6}}
+{\frac{116760}{4463}}\frac{M^{7}}{r^{7}}
-{\frac{126000}{4463}}\frac{M^{8}}{r^{8}} \right)\left( 3 \cos^{2}(\theta) -1 \right),\\
\delta \Psi_3^{\mathrm{GB}}=
&~\sqrt\zeta\chi&{\frac{\sqrt{937230}}{700}}\frac{\Theta f}{{M}^{2}} \frac{M^{4}}{r^{4}}\sin \left( \theta \right)\nonumber\\
&~+\sqrt\zeta\chi^2&{\frac{i37537\sqrt{937230}}{43737400}}{\frac{f }{{M}^{2}\Theta} \frac{M^{5}}{r^{5}}\left(
1
+{\frac{10210}{37537}}\frac{M}{r}
-{\frac{1132390}{112611}}\frac{M^{2}}{r^{2}}
-{\frac{3754100}{112611}}\frac{M^{3}}{r^{3}}
-{\frac{1999200}{37537}}\frac{M^{4}}{r^{4}}
\right) \cos \left( \theta \right)\sin \left( \theta \right) }\nonumber\\
&~-\zeta\chi^2&{\frac{i4463\sqrt{2}}{3500}}\frac{  {f}^{3/2}}{{M}^{2}}\frac{M^{5}}{r^{5}}\left(
1
+{\frac{21440}{4463}}\frac{M}{r}
+{\frac{508960}{31241}}\frac{M^{2}}{r^{2}}
+{\frac{135300}{4463}}\frac{M^{3}}{r^{3}}
+{\frac{167600}{4463}}\frac{M^{4}}{r^{4}}
\right)\sin^{2}(\theta). 
\end{eqnarray}
\end{widetext}

\end{document}